\documentclass[twocolumn]{aastex631}
\usepackage{lineno}

\newcommand{\metallicity}{12 + \log({\rm O/H)}}

\newcommand{\Hg}{H$\gamma$}
\newcommand{\Hb}{H$\beta$}
\newcommand{\EWHb}{EW(H$\beta$)}
\newcommand{\EWHa}{EW(H$\alpha$)}
\newcommand{\Ha}{H$\alpha$}
\newcommand{\oiid}{[\ion{O}{2}]$\lambda\lambda3727,29$}
\newcommand{\oiii}{[\ion{O}{3}]$\lambda4363$}
\newcommand{\oiiid}{[\ion{O}{3}]$\lambda4959,5007$}
\newcommand{\oiiida}{[\ion{O}{3}]$\lambda4959$}
\newcommand{\oiiidb}{[\ion{O}{3}]$\lambda5007$}
\newcommand{\oiidh}{[\ion{O}{2}]$\lambda\lambda7320,30$}
\newcommand{\siid}{[\ion{S}{2}]$\lambda\lambda6717,6731$}

\newcommand{\Hii}{\ion{H}{2}}
\newcommand{\Op}{${\rm O}^{+}$}
\newcommand{\Opp}{${\rm O}^{2+}$}

\newcommand{\toii}{${\rm T}_{\rm e}$(\ion{O}{2})}
\newcommand{\toiii}{${\rm T}_{\rm e}$(\ion{O}{3})}

\newcommand{\jwstDanial}{138}
\newcommand{\jwstDanialNew}{122}
\newcommand{\jwstDanialOii}{20}
\newcommand{\jwstNakajima}{10}
\newcommand{\jwstSanders}{14}
\newcommand{\jwstMorishita}{9}
\newcommand{\jwstLaseter}{10}
\newcommand{\jwstLiterature}{33}
\newcommand{\jwst}{171}
\newcommand{\sdss}{1081}
\newcommand{\sdssOii}{876}
\newcommand{\empress}{103}
\newcommand{\sandersLowz}{17}
\newcommand{\revalski}{12}
\newcommand{\groundBased}{1213}
\newcommand{\sdssStacks}{126}
\newcommand{\sdssStacksOii}{108}
\newcommand{\total}{1510}
\newcommand{\totaloii}{1004}

\newcommand{\oneSigmaOffset}{0.09}

\begin{document}

\title{Genesis-Metallicity: Universal Non-Parametric Gas-Phase Metallicity Estimation}

\correspondingauthor{Danial Langeroodi}
\email{danial.langeroodi@nbi.ku.dk}

\author[0000-0001-5710-8395]{Danial Langeroodi}
\affil{DARK, Niels Bohr Institute, University of Copenhagen, Jagtvej 155A, 2200 Copenhagen, Denmark}

\author[0000-0002-4571-2306]{Jens Hjorth}
\affil{DARK, Niels Bohr Institute, University of Copenhagen, Jagtvej 155A, 2200 Copenhagen, Denmark}

\begin{abstract}

We introduce \texttt{genesis-metallicity}, a gas-phase metallicity measurement \texttt{python} software employing the direct and strong-line methods depending on the available oxygen lines. The non-parametric strong-line estimator is calibrated based on a kernel density estimate in the 4-dimensional space of O2 = \oiid/\Hb; O3 = \oiiidb/\Hb; \Hb\ equivalent width EW(\Hb); and gas-phase metallicity $\metallicity$. We use a calibration sample of \total\ galaxies at $0 < z < 10$ with direct-method metallicity measurements, compiled from the JWST/NIRSpec and ground-based observations. In particular, we report \jwstDanialNew\ new NIRSpec direct-method metallicity measurements at $z > 1$. We show that the O2, O3, and EW(H$\beta$) measurements are sufficient for a gas-phase metallicity estimate that is more accurate than \oneSigmaOffset\ dex. Our calibration is universal, meaning that its accuracy does not depend on the target redshift. Furthermore, the direct-method module employs a non-parametric \toii\ electron temperature estimator based on a kernel density estimate in the 5-dimensional space of O2, O3, \EWHb, \toii, and \toiii. This \toii\ estimator is calibrated based on \totaloii\ spectra with detections of both \oiii\ and \oiidh, notably reporting \jwstDanialOii\ new NIRSpec detections of the \oiidh\ doublet. We make \texttt{genesis-metallicity} and its calibration data publicly available and commit to keeping both up-to-date in light of the incoming data.

\end{abstract}


\section{Introduction} \label{sec: intro}

The ``direct-method'' provides a highly reliable measure of the gas-phase metallicity in galaxies. However, this method relies on an estimate of the electron temperature before the ionic abundances can be derived from the abundance-sensitive emission lines. Unfortunately, the often-faint temperature-sensitive emission lines such as \oiii\ and \oiidh\ remain mostly elusive in large spectroscopic surveys, hindering the application of the direct-method on large samples. This has made the ``strong'' emission lines such as the \oiid\ and \oiiid\ doublets the most commonly used proxies for the gas-phase metallicities of galaxies with available rest-optical spectroscopy \citep[see, e.g.,][]{
2005ApJ...635..260S, 2006ApJ...644..813E, 2008A&A...488..463M, 2009MNRAS.398.1915M, 2011ApJ...730..137Z, 2014ApJ...792...75Z, 2012ApJ...755...73W, 2016ApJ...827...74W, 2013ApJ...772..141B, 2013ApJ...776L..27H, 2013ApJ...774..130K, 2014MNRAS.440.2300C, 2014MNRAS.437.3647Y, 2014ApJ...792....3M, 2014ApJ...795..165S, 2014A&A...563A..58T, 2016ApJ...826L..11K, 2015ApJ...802L..26K, sanders+2015, sanders+2021, 2016MNRAS.463.2002H, 2016ApJ...822...42O, 2017ApJ...849...39S, curti+2017, curti+2024, ZEIGHT, EVOLFMR, 2023NatAs...7.1517H, nakajima+2023, 2024arXiv240717110C, 2024arXiv240807974S}. This practice is commonly known as the ``strong-line'' metallicity estimation: several polynomial relations between various strong-line ratios and gas-phase metallicity are calibrated either empirically on samples with direct-method measurements \cite[see, e.g.,][]{1991ApJ...380..140M, 2010ApJ...720.1738P, 2016MNRAS.457.3678P, curti+2017, 2019ApJ...872..145J, nakajima+2022} or against the predictions of photoionization models \cite[see, e.g.,][]{1985ApJS...57....1M, 2002MNRAS.330...69D, 2002ApJS..142...35K, 2023MNRAS.526.3504H}. 

Despite the success of traditional strong-line metallicity estimators in enabling statistically significant chemical enrichment studies across a wide range of galaxy properties and redshift (see references above), they come with nuanced caveats rooted in their ``parametric'' nature. Firstly; the 2D projections of the calibration data onto the line ratio vs. metallicity planes risk overlooking the complexities of the higher-order parameter space. Even the 2D projections are often too complex to be fully captured by polynomials. For instance, particularly at low metallicities, large scatter is reported around the best-fit O2--$\log(\rm O/H)$ and O32--$\log(\rm O/H)$ relations. \cite{nakajima+2022} showed that the offsets from these best-fit relations depend on the ionization state of interstellar media (ISM), and can be captured by the equivalent width of \Hb, \EWHb. 

Second; the parametric calibrations are prone to ``hot'' spots which render the estimates in certain metallicity windows highly uncertain. For instance, the best-fit polynomials to the O3--$\log(\rm O/H)$ and R23--$\log(\rm O/H)$ projections are widely used as primary metallicity estimators because these relations exhibit relatively tight scatter. However, both projections are non-monotonic, with a turnover metallicity of $\metallicity \sim 8$. This means that i) multiple metallicity solutions exist for each input O3 and R23, which should be sifted based on other projections; ii) the metallicity estimation around the turnover value is highly uncertain due to the flattening of the calibration curve; and iii) observed line ratios higher than the maximum allowed by the calibration curve universally yield the turnover metallicity, failing to capture the intrinsic scatter of the relation. 

Third; recent parametric calibrations at high redshifts based on NIRSpec spectroscopy indicate noticeable deviations from the local-universe calibrations \citep{sanders+2023, laseter+2024}, potentially suggesting a non-universality in the strong-line method. However, as shown by \cite{2018A&A...612A..94N} and \cite{nakajima+2022}, the 2D-projected relationships between the line ratios and gas-phase metallicity are influenced by the ionization parameter. Therefore, high-redshift deviations from the locally-calibrated parametric strong-line estimators are expected, as the high-redshift galaxies exhibit systematically higher ionization parameters. This is evidenced by their observed extremely high O32, \EWHb, and \EWHa\ values \citep{ZEIGHT, langeroodi+2024(stacks), 2023ApJ...952..143R}, indicative of high ionization parameters \citep{2002ApJS..142...35K, 2023MNRAS.526.3504H} and bursty star formation histories \citep{2016ApJ...833..254S, langeroodi+2024(stacks)}. Nonetheless, it is essential for any strong-line calibration to capture such dependencies and remain insensitive to these systematics. 

Here, we overcome these caveats by developing a ``non-parametric'' strong-line metallicity estimator. We achieve this by a kernel density estimate \citep[KDE;][]{1986desd.book.....S, 1992mde..book.....S} of the probability density function (PDF) in the multi-dimensional space of emission line observables and gas-phase metallicity (Section \ref{sec: strong}). This PDF is then used to estimate the gas-phase metallicity for any combination of input emission line observables. We calibrate our strong-line estimator on a sample of \total\ galaxies at $0 < z < 10$ with direct-method metallicity measurements, the largest of such compilations to date (Sections \ref{sec: data} and \ref{sec: direct}). In particular, we report \jwstDanialNew\ new direct-method metallicity measurements at $z > 1$ based on NIRSpec multi-shutter assembly \citep[MSA;][]{jakobsen+2022, ferruit+2022} spectroscopy; this corresponds to a $\sim 6$ fold increase in the sample size of $z > 1$ directly-measured metallicities. We show that the O2, O3, and EW(H$\beta$) measurements are sufficient for a gas-phase metallicity estimate that is more accurate than \oneSigmaOffset\ dex. Our calibration is universal, meaning that its accuracy does not depend on the target redshift. We make \texttt{genesis-metallicity} \citep{genesis_metallicity} and its calibration data available at \url{https://github.com/langeroodi/genesis_metallicity}.

\section{Data} \label{sec: data}

\begin{figure*}
    \centering
    \includegraphics[width=16cm]{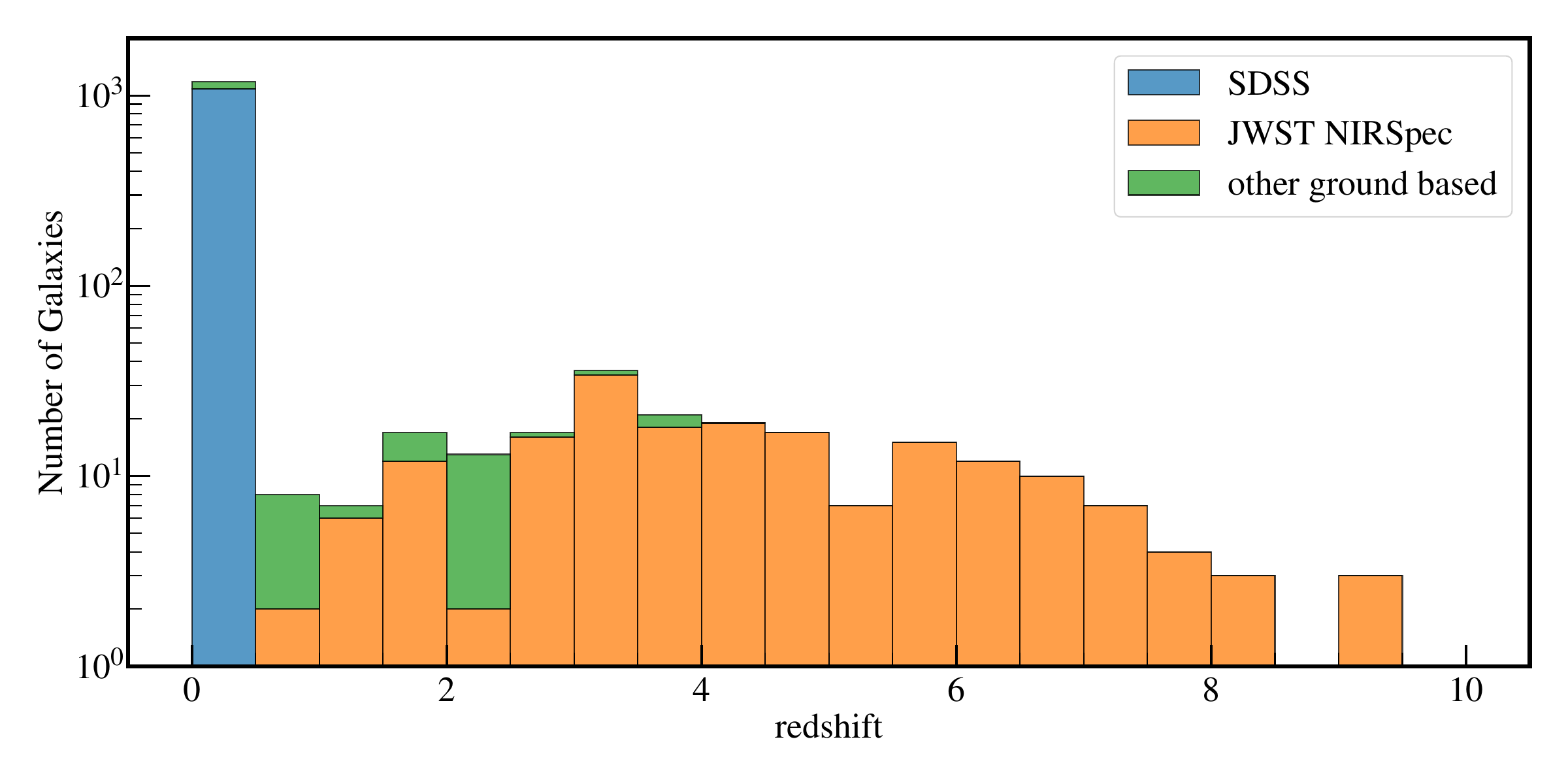}
    \caption{Overview of the calibration sample used in this work. This includes \total\ spectra with direct-method metallicity measurements, including \jwst\ galaxies observed with the NIRSpec MSA (orange), \groundBased\ galaxies observed with ground-based instruments (blue and green), and \sdssStacks\ spectra generated by stacking the SDSS spectra.}
    \label{fig: sample}
\end{figure*}

In this Section, we present an overview of the spectra utilized in our strong-line metallicity calibration. This data consists of \total\ spectra with direct-method metallicity measurements, including \jwst\ galaxies observed with the NIRSpec MSA, \jwstDanialNew\ of which are reported for the first time in this work and the rest are taken from the literature (Section \ref{sec: data: jwst}); \groundBased\ galaxies observed with ground-based instruments (Section \ref{sec: data: ground}); and \sdssStacks\ high-metallicity spectra generated by stacking the SDSS spectra (Section \ref{sec: data: stacks}). Figure \ref{fig: sample} provides an overview of this sample. The line fluxes are reported in Table \ref{table: overview}. We note that the H$\delta$, \Hg, \Hb, and \Ha\ Balmer lines are used for dust reddening correction of emission lines. For this purpose, we assumed a \cite{calzetti+2000} dust curve\footnote{Several recent studies have found evidence that while non-negligible dust is already present in early-universe galaxies, their attenuation curves might deviate from those in local universe \citep{2024arXiv241014671L, 2024arXiv241202557C, 2024arXiv241023959B, 2025arXiv250413118B, 2025NatAs...9..458M, 2025arXiv250412378M, 2025arXiv250214031M, 2025MNRAS.539..109F}. As such, a dedicated investigation of potential systematic biases caused by adopting the \cite{calzetti+2000} or other similar dust curves in this and other similar work is timely. However, this goes beyond the scope of our current study.} and case-B recombination\footnote{The dust attenuation module is available at \url{https://github.com/langeroodi/genesis_metallicity}}.

\subsection{NIRSpec} \label{sec: data: jwst}

We searched the JADES DR3 \citep{jades_dr3} NIRSpec MSA medium-resolution\footnote{Due to its low spectral resolution, the prism grating almost never resolves the \oiii\ line from \Hg. Exceptions can occur at $z > 9$, where these lines fall at relatively high-resolution ($R \sim 300$) prism wavelengths \citep{williams+2023, 2024A&A...687L..11S, 2024arXiv240702575C}.} spectra for \oiii\ and \oiidh\ detections. For this purpose, we used \texttt{pPXF} \citep{pPXF1, pPXF2, pPXF3} to measure the emission line fluxes. For the objects covered in multiple JADES observations, we stacked the spectra from repeated gratings to enhance the signal. We adopted the spectroscopic redshifts reported by the JADES team as a starting point, and for each object ran \texttt{pPXF} on the medium-resolution spectra covering its \oiii\ and \oiidh\ emission. We then visually inspected the subsample with either \oiii\ or \oiidh\ flux signal-to-noise ratios (S/N) greater than 3. We confirm \jwstDanial\ galaxies with robust \oiii\ detections (${\rm S/N} > 3$), \jwstDanialOii\ of which also exhibit robust \oiidh\ detection (${\rm S/N} > 3$). These exclude the confirmed broad-line AGN from \cite{2024A&A...691A.145M}. We also fitted the prism spectra of the \oiii-detected sample to achieve full coverage of the \oiid; \Hg; \oiii; \Hb; \oiiid; \Ha; and \oiidh\ lines. The NIRSpec medium-resolution spectra as well as the corresponding best-fit pPXF models for some example galaxies with \oiii\ and \oiidh\ detections are presented in Appendix \ref{app: detections} and Figures \ref{fig: O4363_1181_00031514_g140m}, \ref{fig: O4363_3215_00265801_g395m}, \ref{fig: O4363_1181_00033391_g235m}, \ref{fig: O4363_3215_00098554_g140m}, \ref{fig: O7320_1180_00013596_g395m}, \ref{fig: O7320_1180_00016375_g395m}, \ref{fig: O7320_1181_00025030_g235m}, and \ref{fig: O7320_1181_00031514_g235m}.

We combined the medium-resolution and prism line flux measurements into a final catalog. We exclusively used the medium resolution measurements for the \Hg, \oiii, and \oiidh\ lines. This is because in prism spectra \Hg\ and \oiii\ lines are rarely deblended from one another and \oiidh\ often appears too faint to be confidently distinguished from the continuum. Since the \Hb\ to \Ha\ flux ratio is the highest-signal Balmer line ratio used to correct for dust attenuation, we prioritized measuring both on the same grating to avoid cross-grating calibration offsets. If multiple gratings provided simultaneous high-significance detections (${\rm S/N} > 3$) of both lines, we prioritized medium-resolution gratings as they generally resolve \Hb\ from \oiiid\ much more comfortably. The \EWHb\ is calculated using the \texttt{pPXF} best-fit continuum on the same grating where the \Hb\ flux is read. For the rest of the lines, if high-significance detections (${\rm S/N} > 3$) are available on the same medium-resolution grating where the \oiii\ line is measured, we prioritize measurements based on this grating. Otherwise, we used the grating that provides the highest S/N flux measurement. We corrected for cross-grating flux calibration offsets by using the brightest line that is covered in both the medium-resolution and prism gratings. At $z < 6$ we avoided the \oiiida\ and \oiiidb\ lines, since they are often blended in prism spectra. Therefore the flux calibration line is often \Hb\ or \Ha. When the medium-resolution flux measurement of a line is adopted, its flux is first normalized by the calibration line flux measured in the same grating, and then multiplied by the calibration line flux measured in the prism grating. 

We also adopted the NIRSpec MSA line fluxes and \EWHb\ measurements for \jwstLiterature\ galaxies from the literature with available direct-method metallicity measurements. This includes \jwstNakajima\ galaxies from \cite{nakajima+2023}, \jwstSanders\ galaxies from \cite{sanders+2023}, and \jwstMorishita\ galaxies from \cite{morishita+2024}. We note that \cite{laseter+2024} reported \jwstLaseter\ galaxies in the JADES DR1 data with direct-method metallicity measurements, which were independently confirmed by the pipeline detailed above. 

\subsection{Ground-based} \label{sec: data: ground}

Our ground-based spectra consists of \sdss\ galaxies selected from the archival SDSS spectra \citep{sdss_dr7}; \empress\ galaxies from the \cite{nakajima+2023} compilation of extremely metal-poor galaxies; \sandersLowz\ galaxies from the \cite{sanders+2020} compilation of $1.5 < z < 3.5$ direct-method metallicity measurements; and \revalski\ galaxies from the MUSE Ultra Deep Field observations \citep{revalski+2024}. Except for the SDSS galaxies, the line fluxes and \EWHb\ for this sample are adopted from the corresponding papers. 

We selected the SDSS galaxies from the MPA-JHU catalog \citep{galSpecLine1, galSpecLine2}. We searched for galaxies where all of the \oiid; \oiii; \Hb; \oiiidb; and \Ha\ lines are detected with ${\rm S/N} > 5$. We sift out the AGNs using the BPT diagram classifications of \cite{galSpecLine2}. We adopted the line fluxes and \EWHb\ as reported in the  MPA-JHU catalog. Because the \oiidh\ flux is not reported in any of the publicly available SDSS catalogs, we used \texttt{pPXF} to measure its flux for the selected galaxies. Out of the \sdss\ selected galaxies, \sdssOii\ galaxies exhibit significant \oiidh\ detections (${\rm S/N} > 3$). 

\subsection{SDSS stacks} \label{sec: data: stacks}

\cite{2013ApJ...765..140A} and \cite{curti+2017} showed that the individual SDSS spectra can be stacked to enhance the \oiii\ signal and enable direct-method metallicity measurements for the less-explored high-metallicity ($8.5 < \metallicity < 9.0$) region of the parameter space. Employing a similar approach, we selected 58207 non-AGN spectra from the MPA-JHU catalog with ${\rm S/N} > 5$ \oiid; \Hb; \oiiidb; and \Ha\ detections. 

We stacked these spectra on a three-dimensional grid of reddening-corrected O2, O3, and \EWHb. This is in contrast with \cite{curti+2017}, where the spectra are stacked on a 2-dimensional grid of reddening-corrected O2 and O3. We chose the 3-dimensional grid because our strong-method calibration relies on O2, O3, and \EWHb\ to estimate the gas-phase metallicity (see Section \ref{sec: strong}). We binned the O2 axis in 0.1 dex intervals, the O3 axis in 0.1 dex intervals, and the \EWHb\ axis in 1 dex intervals. We stacked the spectra using the stacking algorithm detailed in \cite{langeroodi+2024(stacks)}. We used \texttt{pPXF} to measure the line fluxes and \EWHb\ for the stacks. As reported by \cite{curti+2017}, the [\ion{Fe}{2}]$\lambda4360$ line is a common source of systematic offsets in \oiii\ flux measurement of very high metallicity galaxies. To avoid such systematics, we add the [\ion{Fe}{2}]$\lambda4360$ line to the list of emission lines fitted by \texttt{pPXF}. We identified \sdssStacks\ stacks with robust \oiii\ detections (${\rm S/N} > 3$), \sdssStacksOii\ of which also exhibit significant \oiidh\ detections (${\rm S/N} > 3$). 

\section{Direct Measurements} \label{sec: direct}

We measure the ionic oxygen abundances and gas-phase metallicities (O/H) by modelling the emission lines with a 2-zone \Hii\ region \citep{1982A&AS...48..299S, 1992AJ....103.1330G}. This corresponds to a bithermal nebula model, where the low-ionization zone containing species such as \Op\ and the high-ionization zone containing species such as \Opp\ are traced by different temperatures. Assuming an electron density ($n_{\rm e}$), temperature-sensitive line ratios can be used to calculate the electron temperature of each zone. In turn, these temperature measurements allow to derive the ionic abundances of each zone from the abundance-sensitive line fluxes. Where available, we use the \oiid\ and \siid\ lines to estimate the electron densities (adopting the \citealt{2009MNRAS.397..903K} and \citealt{2010ApJS..188...32T} collision strengths, respectively), while assuming $n_{\rm e} = 100\;{\rm cm}^{-3}$ otherwise. The derived temperatures and abundances are only weakly sensitive to the assumed electron density at the density regimes common for galaxies \citep[see, e.g.,][]{curti+2017, nakajima+2023, isobe+2023}. We describe the electron temperature measurements in Section \ref{sec: direct: temperatures} and ionic abundances and gas-phase metallicity measurements in Section \ref{sec: direct: metallicities}. These measurements are reported in Table \ref{table: overview}.  

\subsection{Electron temperatures} \label{sec: direct: temperatures}

\begin{figure*}
    \centering
    \includegraphics[width=14.5cm]{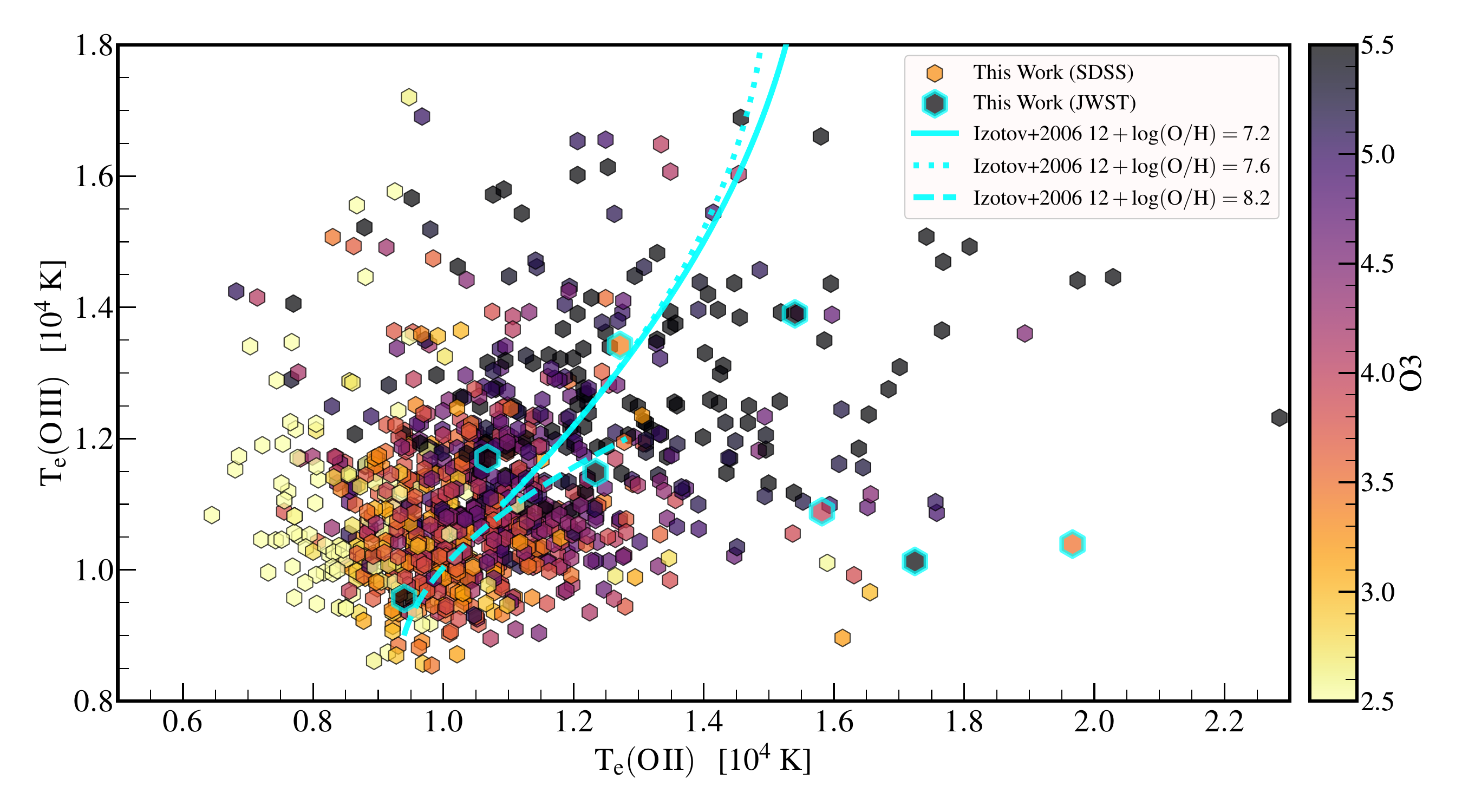}
    \caption{Directly measured \toii\ and \toiii\ for the \totaloii\ spectra where both measurements are available. Each data point is color-coded with its corresponding O3 measurement. The \toii\ and \toiii\ seem to be correlated for average galaxies, with a large scatter that is captured by the O3 value. The x- and y-axis limits are chosen to optimize data visualization; a small number of data points lie outside the displayed range. The full dataset is available in machine-readable format (see Table \ref{table: overview}).}
    \label{fig: temperatures}
\end{figure*}

\begin{figure*}
    \centering
    \includegraphics[width=15cm]{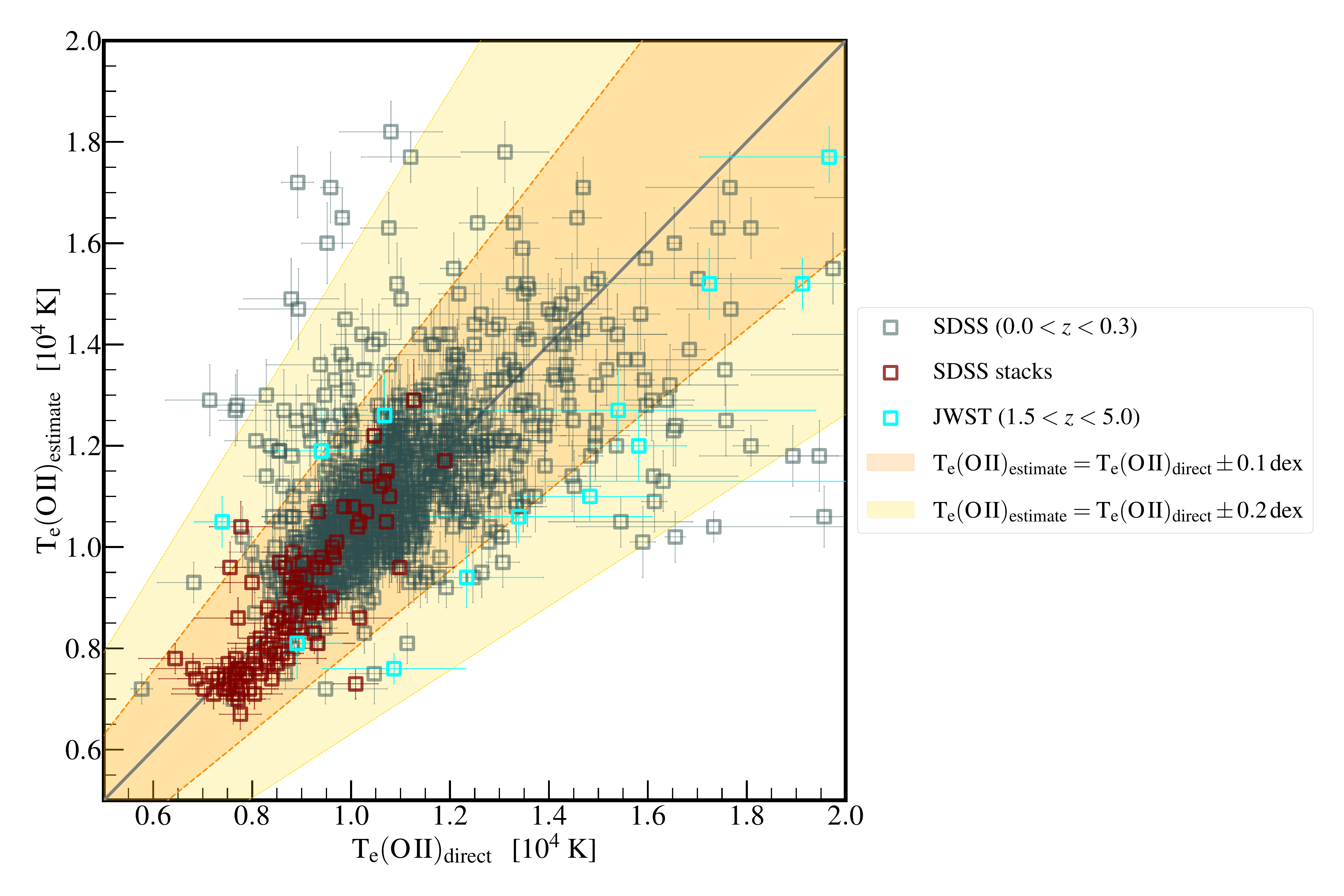}
    \caption{Evaluating the accuracy of the \toii\ estimator. Here, we show the \toii\ estimates vs. the values measured directly from the \oiid/\oiidh\ ratios. The \toii\ is estimated employing a kernel density estimation of the probability density function in the 5-dimensional space of O2, O3, \EWHb, \toii, and \toiii. The \toii\ estimator is more accurate than 0.04 dex, defined as the absolute estimate vs. directly measured \toii\ offset that contains $68\%$ of the estimates. The accuracy of the \toii\ estimator declines to 0.1 dex at \toii\ $> 14000$ K, where the parameter space is sparsely sampled by the calibration data (see Figure \ref{fig: temperatures}).}
    \label{fig: performance t}
\end{figure*}

We measure the \Op\ and \Opp\ electron temperatures, denoted as \toii\ and \toiii, respectively from the \oiid/\oiidh\ and \oiii/\oiiidb\ flux ratios. We used the \texttt{getTemDen} routine of \texttt{PyNeb} \citep{pyneb1, pyneb2} for this purpose, adopting the \cite{2009MNRAS.397..903K} and \cite{2012MNRAS.423L..35P} collision strengths respectively for measuring the \Op\ and \Opp\ electron temperatures\footnote{We note that the measured electron temperatures are somewhat sensitive to the adopted collision strengths table, with reported inconsistencies as high as 500K \citep{2013ApJS..207...21N}.}. This results in \toiii\ measurements for \total\ spectra, \totaloii\ of which also have \toii\ measurements. Figure \ref{fig: temperatures} shows the \toii\ and \toiii\ temperatures for the subsample where both measurements are available.

We find that there is a clear empirical trend between \toii, \toiii, O2, O3, and \EWHb. For instance, the \toii--\toiii--O3 trend is shown in Figure \ref{fig: temperatures}, where the data points are color-coded with their corresponding O3 measurements. Such relations are expected from the photoionization models \citep{izotov+2006}. In particular, a linear relation between \toii\ and \toiii\ is frequently reported in the literature \citep{1986MNRAS.223..811C, 1992AJ....103.1330G, izotov+2006, 2006MNRAS.370.1928P, 2006MNRAS.367.1139P, 2009MNRAS.398..485P, 2010ApJ...720.1738P, curti+2017}, and often proposed for estimating one temperature from the other when needed (i.e., when the required lines are not covered/detected). Although Figure \ref{fig: temperatures} confirms the proposed trends for average galaxies, it also shows considerable scatter around such relations. 

We capture the complex relation between these parameters non-parametrically by employing a kernel density estimate \citep[KDE;][]{1986desd.book.....S, 1992mde..book.....S} in the 5-dimensional space of O2, O3, \EWHb, \toii, and \toiii. The multivariate KDE converts the multi-dimensional distribution of data into a non-parametric estimation of the probability density function (PDF). In turn, this PDF can be used to estimate the probability of specific parameter combinations. We estimate the 5-dimensional PDF using the \texttt{scipy} \citep{scipy} implementation of the \cite{1992mde..book.....S} KDE algorithm with Gaussian kernels (our multivariate probability density estimation is described in more detail in Appendix \ref{app: KDE-bandwidths}; real-valued 5D images of this probability density estimate are available online in machine-readable format). We use the estimated PDF to set up an algorithm, which for each set of input O2, O3, \EWHb, and \toiii\ estimates \toii. This particular configuration is chosen because at high redshifts it is often the case where \oiid; \oiii; \Hb; and \oiiid\ are detected, while \oiidh\ is redshifted out of coverage. As such, it is often the case where measurements of O2, O3, \EWHb, and \toiii\ are available, while \toii\ cannot be directly measured. 

For each set of input O2, O3, \EWHb, \toiii, and their $1\sigma$ uncertainties we make a 4-dimensional grid spanning the $-1\sigma$ to $+1\sigma$ range of each parameter in equally spaced intervals. Assuming that the $-1\sigma$ and $+1\sigma$ uncertainties describe half-Gaussian distributions, we assign a weight to each grid point in this 4-dimensional space. At each grid point, we calculate the probability along the \toii\ axis in 10 K intervals. The resulting 1-dimensional PDFs are multiplied by the weights of the corresponding grid points and then combined to make a 1-dimensional \toii\ PDF. This PDF is used to estimate the best-fit \toii\ and its uncertainty as the highest-probability point and the $1\sigma$ region. 

We evaluate the accuracy of our \toii\ estimator through a leave-one-out cross-validation approach (additional validity tests are provided in Appendix \ref{app: KDE-validity}). In each iteration we take out one data point from the calibration sample, use the KDE on the remaining data points to estimate the 5-dimensional PDF, and apply the resulting \toii\ estimator on the O2, O3, \EWHb, and \toiii\ of the removed data point to estimate its \toii. Figure \ref{fig: performance t} shows the estimated \toii\ vs. the directly measured values. The \toii\ estimator is more accurate than 0.04 dex, defined as the absolute estimate vs. directly measured \toii\ offset that contains $68\%$ of the estimates. As shown in Figure \ref{fig: performance t}, the accuracy of our \toii\ estimator declines to 0.1 dex at \toii\ $>14000$ K, where the parameter space is sparsely sampled (see Figure \ref{fig: temperatures}). Figure \ref{fig: performance t} also shows that the accuracy of our \toii\ estimator declines for high-redshift galaxies. This is mostly driven by the limited coverage of parameter space for these galaxies, where only \jwstDanialOii\ \oiidh\ detections are available.

\subsection{Metallicities} \label{sec: direct: metallicities}

We calculate the \Opp\ ionic abundances from the \oiiid/\Hb\ line ratios, employing the \texttt{getIonAbundance} routine of \texttt{PyNeb} and assuming the \toiii\ electron temperatures calculated in Section \ref{sec: direct: temperatures}. Similarly, the \Op\ ionic abundances are calculated from the \oiid/\Hb\ line ratios, assuming the \toii\ electron temperatures calculated in Section \ref{sec: direct: temperatures}. Whenever there is no \toii\ measurement available, we used the \toii\ estimator calibrated in Section \ref{sec: direct: temperatures} to estimate the \toii\ based on the measured O2, O3, \EWHb, and \toiii. We assume that the \Op\ and \Opp\ are the most abundant oxygen ions, and derive the oxygen abundances (gas-phase metallicity) as the sum of \Op\ and \Opp\ ionic abundances. 

\begin{deluxetable*}{lccccccc}
\tablewidth{0pt}
\tablecaption{Inferred Properties for the Galaxies in the Calibration Sample. Only a small subsample of the calibration data and inferred properties are presented here. The full table containing \total\ galaxies, along with their program and MSA IDs (for JWST sources), SDSS identifiers (for SDSS sources), redshifts, sky coordinates, observed and reddening-corrected emission line fluxes, \Hb\ equivalent widths, and inferred dust attenuation, electron temperatures, and direct-method gas-phase metallicities is available in machine-readable format at the publisher's webpage and at \url{https://github.com/langeroodi/genesis_metallicity}}
\label{table: overview}
\tablehead{
\colhead{ID}
 & \colhead{redshift}
 & \colhead{RA}
 & \colhead{DEC}
 & \colhead{$A_{\rm V}$}
 & \colhead{\toii}
 & \colhead{\toiii}
 & \colhead{metallicity}
 \\ 
\colhead{}
 & \colhead{}
 & \colhead{[deg]}
 & \colhead{[deg]}
 & \colhead{[mag]}
 & \colhead{[$10^4$K]}
 & \colhead{[$10^4$K]}
 & \colhead{$12 + \log({\rm O/H})$}
}
\startdata
1181-00000095 & 3.91 & 189.12919839 & 62.21514056 & 0.59 & $1.496\pm0.230$ & $1.916\pm0.185$ & $7.59\pm0.07$ \\ 
1181-00000902 & 4.06 & 189.19327630 & 62.25372707 & 0.65 & $1.400\pm0.075$ & $1.179\pm0.095$ & $8.14\pm0.09$ \\ 
1181-00000910 & 4.41 & 189.11344299 & 62.25480338 & 1.23 & $1.464\pm0.095$ & $2.147\pm0.336$ & $7.74\pm0.06$ \\ 
1181-00000946 & 4.70 & 189.08587231 & 62.25904450 & 0.01 & $1.350\pm0.324$ & $1.335\pm0.139$ & $7.81\pm0.11$ \\ 
1181-00000956 & 5.42 & 189.10636574 & 62.25971421 & 0.84 & $1.445\pm0.158$ & $1.626\pm0.095$ & $7.83\pm0.05$ \\ 
1181-00000971 & 4.42 & 189.13093211 & 62.26199976 & 0.00 & $1.471\pm0.088$ & $2.122\pm0.350$ & $7.55\pm0.09$ \\ 
1181-00000988 & 6.31 & 189.16214905 & 62.26381085 & 0.72 & $1.496\pm0.156$ & $1.950\pm0.131$ & $7.41\pm0.05$ \\ 
1181-00001048 & 3.87 & 189.05831550 & 62.27255829 & 1.57 & $1.540\pm0.400$ & $1.390\pm0.087$ & $8.03\pm0.07$ \\ 
1181-00001083 & 3.80 & 189.16022669 & 62.27622688 & 0.69 & $1.477\pm0.421$ & $1.740\pm0.285$ & $7.81\pm0.13$ \\ 
1181-00001121 & 3.34 & 189.12982492 & 62.28116924 & 1.46 & $1.011\pm0.102$ & $2.447\pm0.357$ & $8.07\pm0.09$ \\ 
1181-00001129 & 7.09 & 189.17975271 & 62.28238705 & 0.07 & $1.409\pm0.075$ & $1.538\pm0.163$ & $7.91\pm0.10$ \\ 
1181-00001137 & 3.66 & 189.10576599 & 62.28337197 & 1.41 & $0.891\pm0.092$ & $1.837\pm0.098$ & $8.31\pm0.13$ \\ 
1181-00001240 & 3.33 & 189.11737086 & 62.29825830 & 0.91 & $1.492\pm0.171$ & $1.837\pm0.128$ & $7.72\pm0.04$ \\ 
1181-00002000 & 5.66 & 189.17594733 & 62.31153443 & 1.49 & $1.287\pm0.338$ & $2.502\pm0.473$ & $7.83\pm0.05$ \\ 
1181-00002864 & 3.36 & 189.14603821 & 62.25379417 & 2.02 & $1.339\pm0.140$ & $2.424\pm0.227$ & $8.02\pm0.02$ \\ 
1181-00002910 & 4.70 & 189.09764723 & 62.26758235 & 0.99 & $1.152\pm0.217$ & $2.668\pm0.237$ & $7.62\pm0.02$ \\ 
1181-00002916 & 3.66 & 189.10773892 & 62.26952483 & 2.51 & $1.525\pm0.355$ & $2.464\pm0.296$ & $7.48\pm0.09$ \\ 
1181-00003008 & 4.53 & 189.12051573 & 62.30316745 & 1.22 & $1.360\pm0.080$ & $1.318\pm0.104$ & $8.09\pm0.08$ \\ 
1181-00003982 & 7.13 & 189.10941261 & 62.23880148 & NaN & $1.458\pm0.300$ & $2.534\pm0.655$ & $7.17\pm0.06$ \\ 
1181-00004379 & 5.99 & 189.21938693 & 62.23824084 & 0.00 & $1.297\pm0.514$ & $2.488\pm0.737$ & $7.60\pm0.04$ \\ 
1181-00004550 & 3.24 & 189.19247822 & 62.23882485 & 0.92 & $1.360\pm0.070$ & $1.348\pm0.080$ & $8.03\pm0.06$ \\ 
1181-00006476 & 2.98 & 189.16094785 & 62.24473159 & 1.55 & $1.383\pm0.164$ & $2.348\pm0.311$ & $7.69\pm0.04$ \\ 
1181-00007351 & 6.05 & 189.10818294 & 62.24714628 & 0.89 & $1.326\pm0.169$ & $2.445\pm0.262$ & $7.52\pm0.02$ \\ 
1181-00007424 & 7.00 & 189.23290476 & 62.24738144 & 1.28 & $1.357\pm0.139$ & $2.394\pm0.239$ & $7.56\pm0.04$ \\ 
1181-00009104 & 6.82 & 189.24526916 & 62.25252927 & 0.57 & $1.454\pm0.097$ & $2.179\pm0.303$ & $7.59\pm0.06$ \\ 
1181-00010886 & 2.96 & 189.20164917 & 62.25993269 & 0.28 & $1.527\pm0.150$ & $2.345\pm0.133$ & $7.35\pm0.06$ \\ 
1181-00012067 & 4.06 & 189.20745025 & 62.26445323 & 1.05 & $1.423\pm0.107$ & $1.570\pm0.253$ & $7.90\pm0.12$ \\ 
1181-00013041 & 7.09 & 189.20377255 & 62.26842735 & 0.76 & $1.475\pm0.122$ & $2.103\pm0.119$ & $7.50\pm0.04$ \\ 
1181-00015529 & 3.87 & 189.21504396 & 62.27700749 & 1.83 & $0.940\pm0.107$ & $0.957\pm0.097$ & $8.55\pm0.13$ \\ 
1181-00016553 & 4.38 & 189.14360285 & 62.28054547 & 1.45 & $1.234\pm0.156$ & $1.148\pm0.101$ & $8.32\pm0.10$ \\ 
1181-00017997 & 3.32 & 189.18488493 & 62.28493708 & 0.27 & $1.373\pm0.198$ & $2.693\pm0.322$ & $7.25\pm0.26$ \\ 
1181-00018533 & 6.67 & 189.12121255 & 62.28640562 & 0.23 & $1.466\pm0.081$ & $2.139\pm0.299$ & $7.60\pm0.07$ \\ 
1181-00018536 & 6.81 & 189.15531435 & 62.28647145 & 1.01 & $1.520\pm0.356$ & $1.175\pm0.104$ & $8.22\pm0.11$ \\ 
1181-00019715 & 9.31 & 189.13832844 & 62.28986544 & NaN & $1.420\pm0.308$ & $2.611\pm0.575$ & $7.12\pm0.05$ \\ 
1181-00021747 & 3.16 & 189.16868321 & 62.23938764 & 0.78 & $1.492\pm0.398$ & $1.838\pm0.297$ & $7.70\pm0.10$ \\ 
1181-00022737 & 3.07 & 189.08222959 & 62.24503872 & 2.10 & $1.227\pm0.400$ & $2.581\pm0.493$ & $7.94\pm0.11$ \\ 
1181-00024266 & 2.96 & 189.16234694 & 62.25368410 & 0.00 & $1.347\pm0.440$ & $2.733\pm0.670$ & $7.17\pm0.36$ \\ 
1181-00025030 & 1.75 & 189.09503562 & 62.25682574 & 0.00 & $1.967\pm0.261$ & $1.039\pm0.082$ & $8.02\pm0.11$ \\ 
1181-00025351 & 3.13 & 189.17337061 & 62.23041045 & 2.34 & $1.428\pm0.080$ & $2.252\pm0.196$ & $7.62\pm0.02$ \\ 
\enddata
\end{deluxetable*}

\begin{figure*}
    \centering
    \includegraphics[width=18cm]{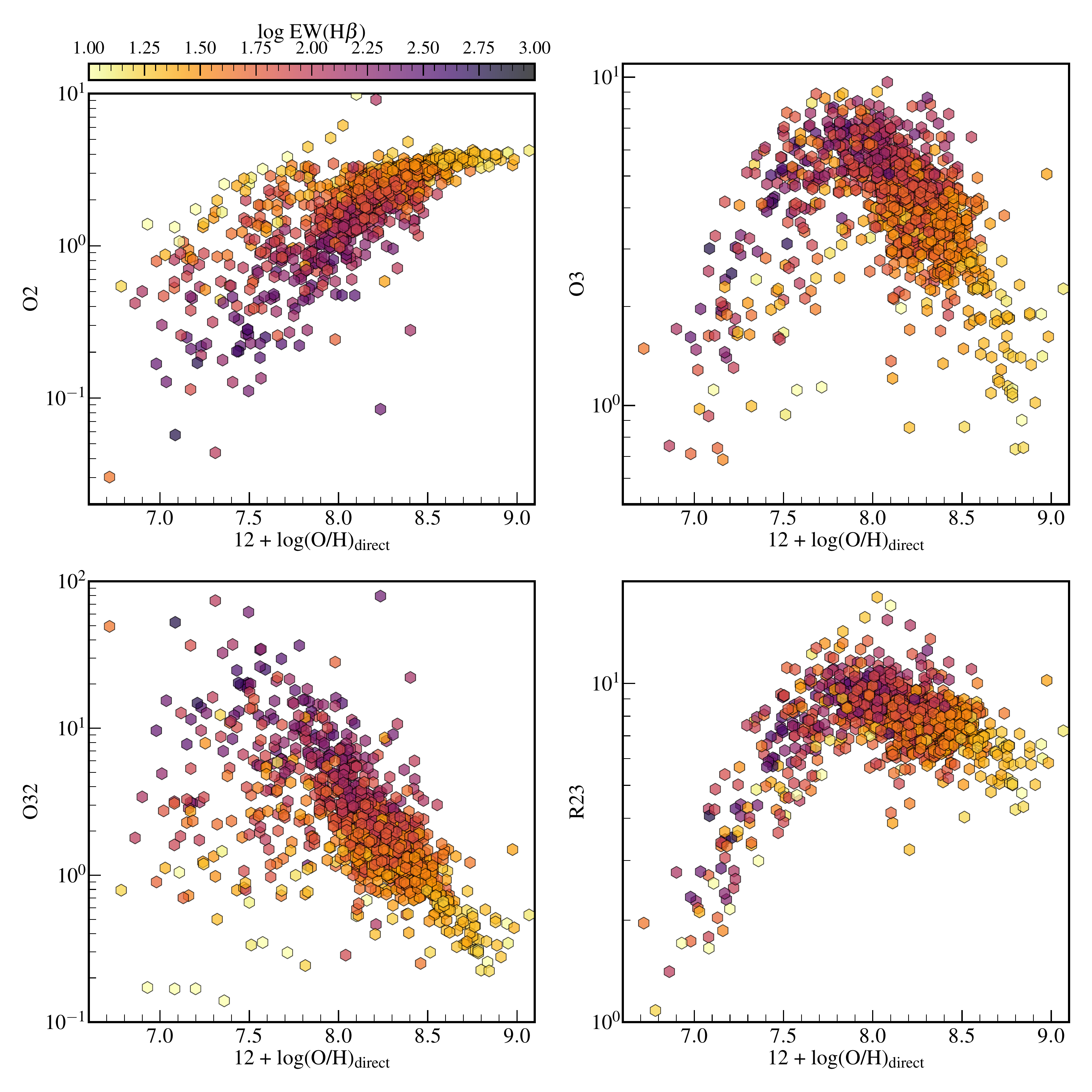}
    \caption{Classic projections of the calibration data onto the 2D planes of O2-$\log(\rm O/H)$, O3-$\log(\rm O/H)$, O32-$\log(\rm O/H)$, and R23-$\log(\rm O/H)$. Each data point is color-coded with its \EWHb\ value. These projections are often used for parametric strong-line metallicity calibrations. We note that these projections are provided here for completeness, and our metallicity estimator is instead calibrated non-parametrically in the 4-dimensional space of O2, O3, \EWHb, and gas-phase metallicity (see Section \ref{sec: strong}) for details.}
    \label{fig: calibration diagrams}
\end{figure*}

\begin{figure*}
    \centering
    \includegraphics[width=15.5cm]{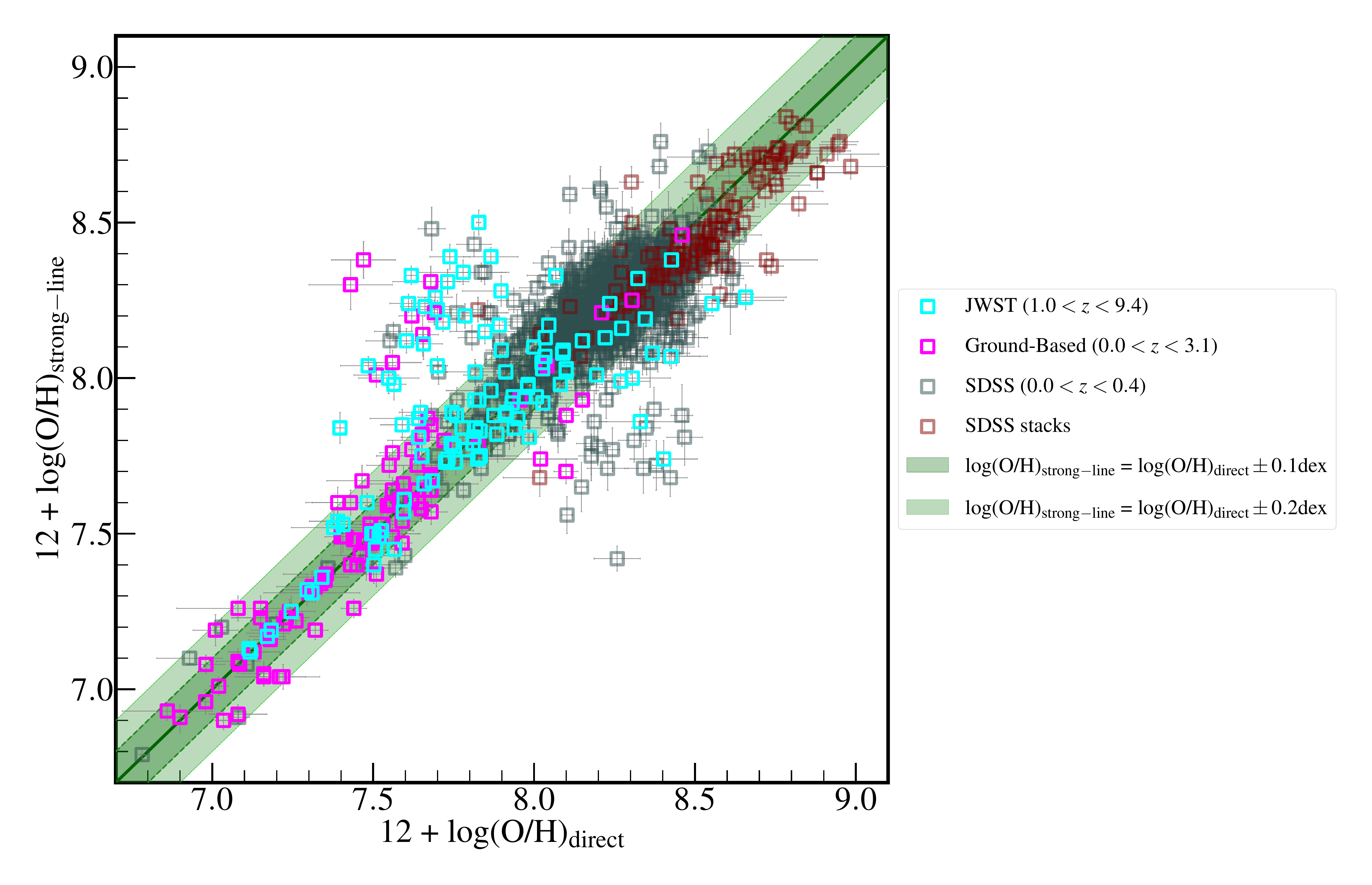}
    \caption{Evaluating the accuracy of the strong-line metallicity estimator. Here, we show the strong-line vs. direct-method metallicities for the calibration sample. This plot is generated through a leave-one-out cross-validation approach. In brief, in each iteration we exclude one data point from the calibration sample, calibrate the metallicity estimator on the remaining sample, and use this estimator to estimate the metallicity of the excluded point based on its O2, O3, and \EWHb. The strong-line metallicity estimator is more accurate than \oneSigmaOffset\ dex, defined as the absolute strong-line vs. direct metallicity offset containing $68\%$ of the estimates. Our calibration is universal, meaning that its accuracy does not depend on the target redshift.}
    \label{fig: strong vs direct}
\end{figure*}

\begin{figure*}
    \centering
    \includegraphics[width=16cm]{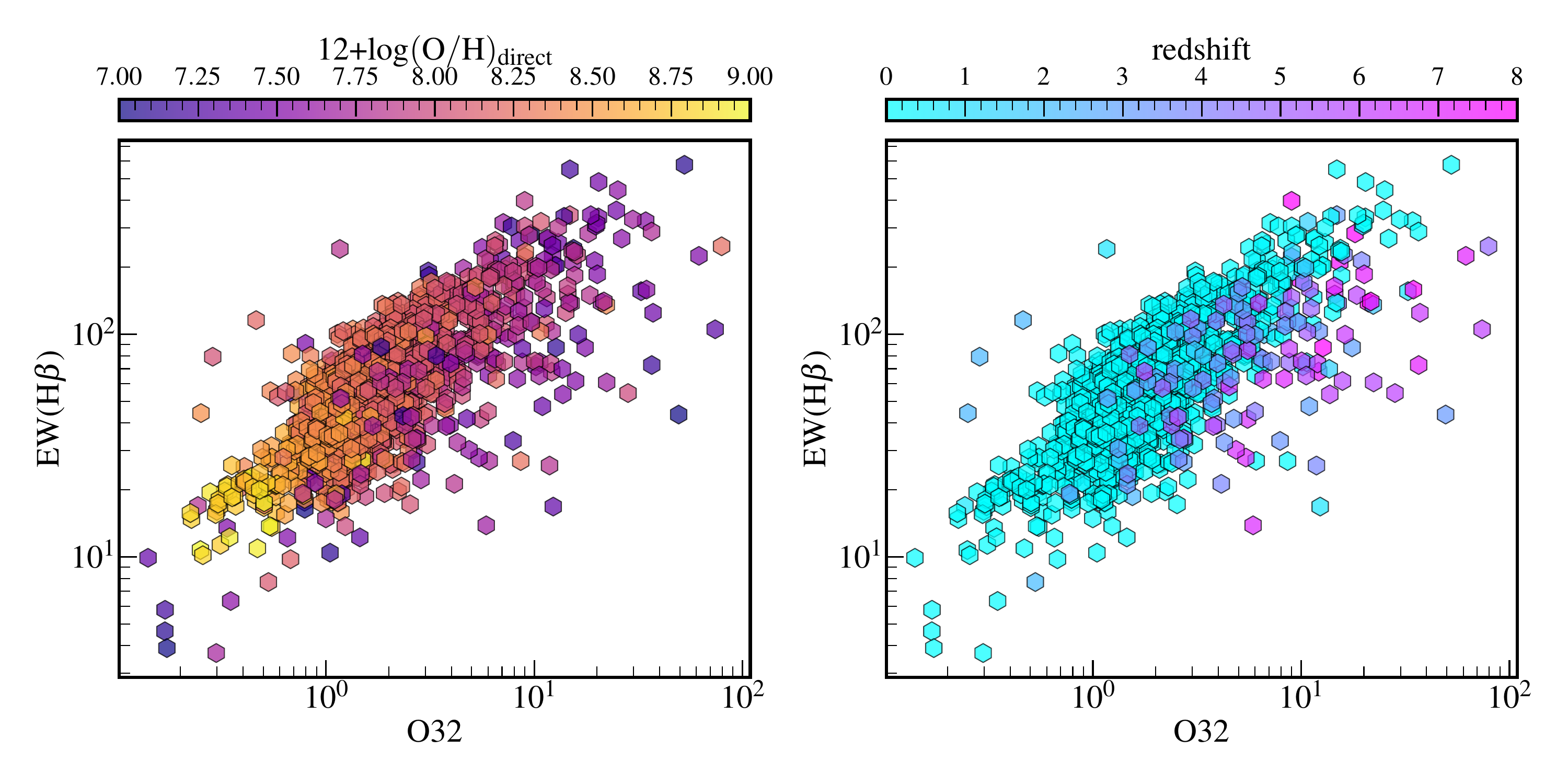}
    \caption{Tight correlation between the O32 line ratio and \EWHb. Since the O32 is an ionization parameter estimator, this correlation suggests that \EWHb\ can also be used to trace the ionization parameter. In the left panel the data points are color-coded with their corresponding directly-measured metallicities. In the right panel the data points are color-coded with their corresponding redshifts.}
    \label{fig: ionization}
\end{figure*}

\section{Strong-line calibration} \label{sec: strong}

We use the distribution of the calibration data in the 4-dimensional space of O2, O3, \EWHb, and gas-phase metallicity for a non-parametric calibration of a strong-line metallicity estimator. Figure \ref{fig: calibration diagrams} shows 4 classic projections of the data, frequently used for the parametric calibration of the strong-line metallicity estimators. We adapt a method similar to that described in Section \ref{sec: direct: temperatures} for the non-parametric calibration. In brief, we use a kernel density estimate (KDE) in the 4-dimensional space of O2, O3, \EWHb, and gas-phase metallicity to estimate the probability density function (PDF) non-parametrically based on the distribution of the calibration data (our multivariate probability density estimation is described in more detail in Appendix \ref{app: KDE-bandwidths}; real-valued 4D images of this probability density estimate are available online in machine-readable format). This PDF is then used to estimate the gas-phase metallicity for any combination of input O2, O3, and \EWHb. The marginalization procedure is described in detail in Section \ref{sec: direct: temperatures}. In this calibration, we only include the subsample with direct-method metallicity uncertainties lower than 0.2 dex.

We evaluate the accuracy of our strong-line metallicity estimator with a leave-one-out cross-validation approach, similar to that described in Section \ref{sec: direct: temperatures} (additional validity tests are provided in Appendix \ref{app: KDE-validity}). In each iteration, we exclude one data point from the calibration sample, calibrate the metallicity estimator on the remaining data, and use this estimator to estimate the metallicity of the excluded point based on its O2, O3, and \EWHb\ measurements. Figure \ref{fig: strong vs direct} shows the strong-line gas-phase metallicity estimates vs. those measured by the direct method. Our metallicity estimations are more accurate than \oneSigmaOffset\ dex, defined as the absolute strong-line vs. direct metallicity offset which contains $68\%$ of the estimates. 

The accuracy of our strong-line metallicity estimator does not vary noticeably with redshift. We achieve a 0.09 dex accuracy at $z < 0.5$, 0.11 dex accuracy at $z > 0.5$, and a 0.12 dex accuracy at $z > 1.0$. We further confirm this by adding the source redshift as an extra dimension to the kernel density estimate and re-calibrating the strong-line metallicity estimator. Repeating the same leave-one-out cross-validation test as above, we achieve identical accuracies at $z < 0.5$, $z > 0.5$, and $z > 1.0$. This highlights that adding the redshift provides no additional information for estimating the gas-phase metallicities beyond what is already captured by O2, O3, and \EWHb. 

As shown in Figure \ref{fig: ionization}, \EWHb\ and O32 are tightly correlated. Since O32 is widely accepted as an ionization parameter estimator \citep[see, e.g.,][]{2002ApJS..142...35K, 2023MNRAS.526.3504H}, this correlation suggests that \EWHb\ closely traces the ionization parameter as well. Photoionization models imply that the relationship between the strong-line ratios and gas-phase metallicity is sensitive to the ionization parameter \citep{2018A&A...612A..94N, nakajima+2022}. As such, including an ionization parameter estimator such as O32 or \EWHb\ in the strong-line metallicity calibration is expected to increase its accuracy. Indeed, \cite{nakajima+2022} showed that the parametric strong-line metallicity calibration can be improved, particularly at the low-metallicity end, by splitting calibration into three ionization branches as traced by the observed \EWHb. 

The tight correlation between the \EWHb\ and O32 (see Figure \ref{fig: ionization}) might suggest that O2 and O3 should be sufficient to calibrate an optimal strong-line metallicity estimator; i.e., suggesting that including the \EWHb\ information is unnecessary. This is because \EWHb\ is accurately predicted from its tight correlation with O32, and the O32 information is already captured by including the O2 and O3. We test this by removing the \EWHb\ axis from our strong-line metallicity calibration, and using a KDE in the 3-dimensional space of O2, O3, and gas-phase metallicity. This slightly yet noticeably decreases the accuracy of our metallicity estimator to 0.17 dex. Hence, \EWHb\ is providing additional information beyond what is captured by O2 and O3. This seems intuitive from Figure \ref{fig: ionization}, where the offset from the average \EWHb-O3 relation seems to correlate with both the gas-phase metallicity and redshift, as indicated by the color-coding in the left and right panels, respectively. Similarly, we find that including the \EWHb\ information slightly improves the accuracy of our \toii\ estimator (Section \ref{sec: direct: temperatures}). 

Nonetheless, we make both our \EWHb-independent electron temperature and metallicity estimators publicly available. A strong-line metallicity calibration that does not rely on \EWHb\ measurements is often required, despite the discussed loss in accuracy when \EWHb\ information is not included. Most importantly, equivalent width measurements are contingent upon continuum detection, which is not always possible particularly for faint high-redshift galaxies. Moreover, equivalent width measurements based on NIRSpec MSA spectra can be subject to several systematic uncertainties as a result of either generic path-loss corrections (treating extended galaxies as point sources or uniformly extended sources) or mismatch between the emission lines and continuum profiles \citep[e.g.,][]{ferruit+2022, jades_dr3}. These systematics are especially pronounced for massive intermediate-redshift galaxies, with resolved morphologies often filling the MSA shutters or even extending beyond them.

Similarly, it might be desirable to avoid explicitly using \EWHb\ in measuring gas-phase metallicities when investigating galaxy scaling relations such as the mass-metallicity and fundamental metallicity relations. This is because \EWHb\ is primarily recognized as a specific SFR (sSFR) tracer, with the \EWHb\ vs. ionization parameter correlation discussed above likely a secondary trend driven by the correlation between sSFR and ionization parameter \citep{2018MNRAS.477.5568K, 2022ApJ...937...22P, 2023ApJ...952..167R}. As such, it might be preferred to use the \EWHb-independent estimator for measuring gas-phase metallicities to avoid introducing sSFR-driven systematic biases in the inferred scaling relations. We argue that this is not a major concern since \EWHb\ (and therefore sSFR by proxy) is accurately traced by O32 (as shown in Figure \ref{fig: ionization}), which is included in our and most other calibrations. Therefore, we advise the user to include the \EWHb\ information where available. However, we note that fully exploring this and other potential systematics which affect scaling relations goes beyond the scope of this study.

\section{Conclusion} \label{sec: con}

We present \texttt{genesis-metallicity}, a non-parametric electron temperature and gas-phase metallicity estimator. This code is calibrated on a sample of \total\ \oiii\ detections at $0 < z < 10$, compiled from the JWST/NIRSpec and ground-based observations. In particular, we report \jwstDanialNew\ new NIRSpec direct-method metallicity measurements at $z > 1$; this corresponds to a $\sim 6$ fold increase in the sample size of $z > 1$ directly-measured metallicities. 

The electron temperature estimator is calibrated based on a kernel density estimate of the probability density function in the 5-dimensional space of O2, O3, \EWHb, \toii, and \toiii. We achieve a 0.04 dex accuracy in our \toii\ estimates. The strong-line metallicity estimator is calibrated in the 4-dimensional space of O2, O3, \EWHb, and gas-phase metallicity. We achieve a 0.09 dex accuracy in our strong-line gas-phase metallicity estimates. Our calibration is universal, meaning that its accuracy does not depend on the target redshift. 

Improved sampling of the sparsely populated regions of the emission line observables parameter space can further enhance the accuracy of our calibration. Therefore, we commit to keeping \texttt{genesis-metallicity} and its calibration data up-to-date in light of the upcoming data. The most recent version of \texttt{genesis-metallicity} and its calibration data can be found at \url{https://github.com/langeroodi/genesis_metallicity}. The version corresponding to this draft (1.2.0) is archived at Zenodo and can be found at \cite{genesis_metallicity}.

\section*{Acknowledgments}

We greatly appreciate comments from the referee on both the manuscript and the public code, which substantially improved the presentation of the data and methodology, as well as the accessibility of the software. We are grateful to the statistics editor, whose feedback helped shape the discussion in Appendix \ref{app: KDE}. We also appreciate comments from Chris Willott, which were crucial in identifying a bug in the initial version (v1.0) of the public code. This work was made possible by the public release of the reduced JWST NIRSpec MSA spectra acquired through the JADES and JOF programs; all the JWST data used in this work can be found in MAST: \dataset[10.17909/8tdj-8n28]{http://dx.doi.org/10.17909/8tdj-8n28}. Moreover, we heavily used the latest release of the SDSS data (DR18). This work was supported by research grants (VIL16599, VIL54489) from VILLUM FONDEN.

\clearpage
\bibliography{main}

@ARTICLE{jades_dr3,
       author = {{D'Eugenio}, Francesco and {Cameron}, Alex J. and {Scholtz}, Jan and {Carniani}, Stefano and {Willott}, Chris J. and {Curtis-Lake}, Emma and {Bunker}, Andrew J. and {Parlanti}, Eleonora and {Maiolino}, Roberto and {Willmer}, Christopher N.~A. and {Jakobsen}, Peter and {Robertson}, Brant E. and {Johnson}, Benjamin D. and {Tacchella}, Sandro and {Cargile}, Phillip A. and {Rawle}, Tim and {Arribas}, Santiago and {Chevallard}, Jacopo and {Curti}, Mirko and {Egami}, Eiichi and {Eisenstein}, Daniel J. and {Kumari}, Nimisha and {Looser}, Tobias J. and {Rieke}, Marcia J. and {Rodr{\'\i}guez Del Pino}, Bruno and {Saxena}, Aayush and {{\"U}bler}, Hannah and {Venturi}, Giacomo and {Witstok}, Joris and {Baker}, William M. and {Bhatawdekar}, Rachana and {Bonaventura}, Nina and {Boyett}, Kristan and {Charlot}, St{\'e}phane and {Danhaive}, A. Lola and {Hainline}, Kevin N. and {Hausen}, Ryan and {Helton}, Jakob M. and {Ji}, Xihan and {Ji}, Zhiyuan and {Jones}, Gareth C. and {Joud{\v{z}}balis}, Ignas and {Maseda}, Michael V. and {P{\'e}rez-Gonz{\'a}lez}, Pablo G. and {Perna}, Michele and {Pusk{\'a}s}, D{\'a}vid and {Shivaei}, Irene and {Silcock}, Maddie S. and {Simmonds}, Charlotte and {Smit}, Renske and {Sun}, Fengwu and {Villanueva}, Natalia C. and {Williams}, Christina C. and {Zhu}, Yongda},
        title = "{JADES Data Release 3 -- NIRSpec/MSA spectroscopy for 4,000 galaxies in the GOODS fields}",
      journal = {arXiv e-prints},
         year = 2024,
        month = apr,
          eid = {arXiv:2404.06531},
        pages = {arXiv:2404.06531},
          doi = {10.48550/arXiv.2404.06531},
archivePrefix = {arXiv},
       eprint = {2404.06531},
 primaryClass = {astro-ph.GA},
       adsurl = {https://ui.adsabs.harvard.edu/abs/2024arXiv240406531D}
}

@ARTICLE{pPXF1,
       author = {{Cappellari}, Michele and {Emsellem}, Eric},
        title = "{Parametric Recovery of Line-of-Sight Velocity Distributions from Absorption-Line Spectra of Galaxies via Penalized Likelihood}",
      journal = {\pasp},
         year = 2004,
        month = feb,
       volume = {116},
       number = {816},
        pages = {138-147},
          doi = {10.1086/381875},
archivePrefix = {arXiv},
       eprint = {astro-ph/0312201},
 primaryClass = {astro-ph},
       adsurl = {https://ui.adsabs.harvard.edu/abs/2004PASP..116..138C}
}

@ARTICLE{pPXF2,
       author = {{Cappellari}, Michele},
        title = "{Improving the full spectrum fitting method: accurate convolution with Gauss-Hermite functions}",
      journal = {\mnras},
         year = 2017,
        month = apr,
       volume = {466},
       number = {1},
        pages = {798-811},
          doi = {10.1093/mnras/stw3020},
archivePrefix = {arXiv},
       eprint = {1607.08538},
 primaryClass = {astro-ph.GA},
       adsurl = {https://ui.adsabs.harvard.edu/abs/2017MNRAS.466..798C}
}

@ARTICLE{pPXF3,
    author = {{Cappellari}, M.},
    title = "{Full spectrum fitting with photometry in ppxf: non-parametric
        star formation history, metallicity and the quenching boundary from
        3200 LEGA-C galaxies at redshift $z\approx0.8$}",
    journal = {MNRAS submitted},
    eprint = {2208.14974},
    year = 2022,
    doi = {10.48550/arXiv.2208.14974}
}

@ARTICLE{2024A&A...687L..11S,
       author = {{Schaerer}, D. and {Marques-Chaves}, R. and {Xiao}, M. and {Korber}, D.},
        title = "{Discovery of a new N-emitter in the epoch of reionization}",
      journal = {\aap},
         year = 2024,
        month = jul,
       volume = {687},
          eid = {L11},
        pages = {L11},
          doi = {10.1051/0004-6361/202450721},
archivePrefix = {arXiv},
       eprint = {2406.08408},
 primaryClass = {astro-ph.GA},
       adsurl = {https://ui.adsabs.harvard.edu/abs/2024A&A...687L..11S}
}

@ARTICLE{2024arXiv240702575C,
       author = {{Curti}, Mirko and {Witstok}, Joris and {Jakobsen}, Peter and {Kobayashi}, Chiaki and {Curtis-Lake}, Emma and {Hainline}, Kevin and {Ji}, Xihan and {D'Eugenio}, Francesco and {Chevallard}, Jacopo and {Maiolino}, Roberto and {Scholtz}, Jan and {Carniani}, Stefano and {Arribas}, Santiago and {Baker}, William M. and {Bhatawdekar}, Rachana and {Boyett}, Kristan and {Bunker}, Andrew J. and {Cameron}, Alex and {Cargile}, Phillip A. and {Charlot}, Stephane and {Eisenstein}, Daniel J. and {Ji}, Zhiyuan and {Johnson}, Benjamin D. and {Kumari}, Nimisha and {Maseda}, Michael V. and {Robertson}, Brant and {Silcock}, Maddie S. and {Tacchella}, Sandro and {Ubler}, Hannah and {Venturi}, Giacomo and {Williams}, Christina C. and {Willmer}, Christopher N.~A. and {Willott}, Chris},
        title = "{JADES: The star-formation and chemical enrichment history of a luminous galaxy at z\raisebox{-0.5ex}\textasciitilde9.43 probed by ultra-deep JWST/NIRSpec spectroscopy}",
      journal = {arXiv e-prints},
         year = 2024,
        month = jul,
          eid = {arXiv:2407.02575},
        pages = {arXiv:2407.02575},
          doi = {10.48550/arXiv.2407.02575},
archivePrefix = {arXiv},
       eprint = {2407.02575},
 primaryClass = {astro-ph.GA},
       adsurl = {https://ui.adsabs.harvard.edu/abs/2024arXiv240702575C}
}

@ARTICLE{williams+2023,
       author = {{Williams}, Hayley and {Kelly}, Patrick L. and {Chen}, Wenlei and {Brammer}, Gabriel and {Zitrin}, Adi and {Treu}, Tommaso and {Scarlata}, Claudia and {Koekemoer}, Anton M. and {Oguri}, Masamune and {Lin}, Yu-Heng and {Diego}, Jose M. and {Nonino}, Mario and {Hjorth}, Jens and {Langeroodi}, Danial and {Broadhurst}, Tom and {Rogers}, Noah and {Perez-Fournon}, Ismael and {Foley}, Ryan J. and {Jha}, Saurabh and {Filippenko}, Alexei V. and {Strolger}, Lou and {Pierel}, Justin and {Poidevin}, Frederick and {Yang}, Lilan},
        title = "{A magnified compact galaxy at redshift 9.51 with strong nebular emission lines}",
      journal = {Science},
         year = 2023,
        month = apr,
       volume = {380},
       number = {6643},
        pages = {416-420},
          doi = {10.1126/science.adf5307},
archivePrefix = {arXiv},
       eprint = {2210.15699},
 primaryClass = {astro-ph.GA},
       adsurl = {https://ui.adsabs.harvard.edu/abs/2023Sci...380..416W}
}

@ARTICLE{nakajima+2023,
       author = {{Nakajima}, Kimihiko and {Ouchi}, Masami and {Isobe}, Yuki and {Harikane}, Yuichi and {Zhang}, Yechi and {Ono}, Yoshiaki and {Umeda}, Hiroya and {Oguri}, Masamune},
        title = "{JWST Census for the Mass-Metallicity Star Formation Relations at z = 4-10 with Self-consistent Flux Calibration and Proper Metallicity Calibrators}",
      journal = {\apjs},
         year = 2023,
        month = dec,
       volume = {269},
       number = {2},
          eid = {33},
        pages = {33},
          doi = {10.3847/1538-4365/acd556},
archivePrefix = {arXiv},
       eprint = {2301.12825},
 primaryClass = {astro-ph.GA},
       adsurl = {https://ui.adsabs.harvard.edu/abs/2023ApJS..269...33N}
}

@ARTICLE{morishita+2024,
       author = {{Morishita}, Takahiro and {Stiavelli}, Massimo and {Grillo}, Claudio and {Rosati}, Piero and {Schuldt}, Stefan and {Trenti}, Michele and {Bergamini}, Pietro and {Boyett}, Kristan N. and {Chary}, Ranga-Ram and {Leethochawalit}, Nicha and {Roberts-Borsani}, Guido and {Treu}, Tommaso and {Vanzella}, Eros},
        title = "{Diverse Oxygen Abundance in Early Galaxies Unveiled by Auroral Line Analysis with JWST}",
      journal = {arXiv e-prints},
         year = 2024,
        month = feb,
          eid = {arXiv:2402.14084},
        pages = {arXiv:2402.14084},
          doi = {10.48550/arXiv.2402.14084},
archivePrefix = {arXiv},
       eprint = {2402.14084},
 primaryClass = {astro-ph.GA},
       adsurl = {https://ui.adsabs.harvard.edu/abs/2024arXiv240214084M}
}

@ARTICLE{sanders+2023,
       author = {{Sanders}, Ryan L. and {Shapley}, Alice E. and {Topping}, Michael W. and {Reddy}, Naveen A. and {Brammer}, Gabriel B.},
        title = "{Direct T $_{e}$-based Metallicities of z = 2{\textendash}9 Galaxies with JWST/NIRSpec: Empirical Metallicity Calibrations Applicable from Reionization to Cosmic Noon}",
      journal = {\apj},
         year = 2024,
        month = feb,
       volume = {962},
       number = {1},
          eid = {24},
        pages = {24},
          doi = {10.3847/1538-4357/ad15fc},
archivePrefix = {arXiv},
       eprint = {2303.08149},
 primaryClass = {astro-ph.GA},
       adsurl = {https://ui.adsabs.harvard.edu/abs/2024ApJ...962...24S}
}

@ARTICLE{laseter+2024,
       author = {{Laseter}, Isaac H. and {Maseda}, Michael V. and {Curti}, Mirko and {Maiolino}, Roberto and {D'Eugenio}, Francesco and {Cameron}, Alex J. and {Looser}, Tobias J. and {Arribas}, Santiago and {Baker}, William M. and {Bhatawdekar}, Rachana and {Boyett}, Kristan and {Bunker}, Andrew J. and {Carniani}, Stefano and {Charlot}, Stephane and {Chevallard}, Jacopo and {Curtis-lake}, Emma and {Egami}, Eiichi and {Eisenstein}, Daniel J. and {Hainline}, Kevin and {Hausen}, Ryan and {Ji}, Zhiyuan and {Kumari}, Nimisha and {Perna}, Michele and {Rawle}, Tim and {Rix}, Hans-Walter and {Robertson}, Brant and {Rodr{\'\i}guez Del Pino}, Bruno and {Sandles}, Lester and {Scholtz}, Jan and {Smit}, Renske and {Tacchella}, Sandro and {{\"U}bler}, Hannah and {Williams}, Christina C. and {Willott}, Chris and {Witstok}, Joris},
        title = "{JADES: Detecting [OIII]{\ensuremath{\lambda}}4363 emitters and testing strong line calibrations in the high-z Universe with ultra-deep JWST/NIRSpec spectroscopy up to z {\ensuremath{\sim}} 9.5}",
      journal = {\aap},
         year = 2024,
        month = jan,
       volume = {681},
          eid = {A70},
        pages = {A70},
          doi = {10.1051/0004-6361/202347133},
archivePrefix = {arXiv},
       eprint = {2306.03120},
 primaryClass = {astro-ph.GA},
       adsurl = {https://ui.adsabs.harvard.edu/abs/2024A&A...681A..70L}
}

@ARTICLE{sdss_dr7,
       author = {{Abazajian}, Kevork N. and {Adelman-McCarthy}, Jennifer K. and {Ag{\"u}eros}, Marcel A. and {Allam}, Sahar S. and {Allende Prieto}, Carlos and {An}, Deokkeun and {Anderson}, Kurt S.~J. and {Anderson}, Scott F. and {Annis}, James and {Bahcall}, Neta A. and {Bailer-Jones}, C.~A.~L. and {Barentine}, J.~C. and {Bassett}, Bruce A. and {Becker}, Andrew C. and {Beers}, Timothy C. and {Bell}, Eric F. and {Belokurov}, Vasily and {Berlind}, Andreas A. and {Berman}, Eileen F. and {Bernardi}, Mariangela and {Bickerton}, Steven J. and {Bizyaev}, Dmitry and {Blakeslee}, John P. and {Blanton}, Michael R. and {Bochanski}, John J. and {Boroski}, William N. and {Brewington}, Howard J. and {Brinchmann}, Jarle and {Brinkmann}, J. and {Brunner}, Robert J. and {Budav{\'a}ri}, Tam{\'a}s and {Carey}, Larry N. and {Carliles}, Samuel and {Carr}, Michael A. and {Castander}, Francisco J. and {Cinabro}, David and {Connolly}, A.~J. and {Csabai}, Istv{\'a}n and {Cunha}, Carlos E. and {Czarapata}, Paul C. and {Davenport}, James R.~A. and {de Haas}, Ernst and {Dilday}, Ben and {Doi}, Mamoru and {Eisenstein}, Daniel J. and {Evans}, Michael L. and {Evans}, N.~W. and {Fan}, Xiaohui and {Friedman}, Scott D. and {Frieman}, Joshua A. and {Fukugita}, Masataka and {G{\"a}nsicke}, Boris T. and {Gates}, Evalyn and {Gillespie}, Bruce and {Gilmore}, G. and {Gonzalez}, Belinda and {Gonzalez}, Carlos F. and {Grebel}, Eva K. and {Gunn}, James E. and {Gy{\"o}ry}, Zsuzsanna and {Hall}, Patrick B. and {Harding}, Paul and {Harris}, Frederick H. and {Harvanek}, Michael and {Hawley}, Suzanne L. and {Hayes}, Jeffrey J.~E. and {Heckman}, Timothy M. and {Hendry}, John S. and {Hennessy}, Gregory S. and {Hindsley}, Robert B. and {Hoblitt}, J. and {Hogan}, Craig J. and {Hogg}, David W. and {Holtzman}, Jon A. and {Hyde}, Joseph B. and {Ichikawa}, Shin-ichi and {Ichikawa}, Takashi and {Im}, Myungshin and {Ivezi{\'c}}, {\v{Z}}eljko and {Jester}, Sebastian and {Jiang}, Linhua and {Johnson}, Jennifer A. and {Jorgensen}, Anders M. and {Juri{\'c}}, Mario and {Kent}, Stephen M. and {Kessler}, R. and {Kleinman}, S.~J. and {Knapp}, G.~R. and {Konishi}, Kohki and {Kron}, Richard G. and {Krzesinski}, Jurek and {Kuropatkin}, Nikolay and {Lampeitl}, Hubert and {Lebedeva}, Svetlana and {Lee}, Myung Gyoon and {Lee}, Young Sun and {French Leger}, R. and {L{\'e}pine}, S{\'e}bastien and {Li}, Nolan and {Lima}, Marcos and {Lin}, Huan and {Long}, Daniel C. and {Loomis}, Craig P. and {Loveday}, Jon and {Lupton}, Robert H. and {Magnier}, Eugene and {Malanushenko}, Olena and {Malanushenko}, Viktor and {Mandelbaum}, Rachel and {Margon}, Bruce and {Marriner}, John P. and {Mart{\'\i}nez-Delgado}, David and {Matsubara}, Takahiko and {McGehee}, Peregrine M. and {McKay}, Timothy A. and {Meiksin}, Avery and {Morrison}, Heather L. and {Mullally}, Fergal and {Munn}, Jeffrey A. and {Murphy}, Tara and {Nash}, Thomas and {Nebot}, Ada and {Neilsen}, Eric H., Jr. and {Newberg}, Heidi Jo and {Newman}, Peter R. and {Nichol}, Robert C. and {Nicinski}, Tom and {Nieto-Santisteban}, Maria and {Nitta}, Atsuko and {Okamura}, Sadanori and {Oravetz}, Daniel J. and {Ostriker}, Jeremiah P. and {Owen}, Russell and {Padmanabhan}, Nikhil and {Pan}, Kaike and {Park}, Changbom and {Pauls}, George and {Peoples}, John, Jr. and {Percival}, Will J. and {Pier}, Jeffrey R. and {Pope}, Adrian C. and {Pourbaix}, Dimitri and {Price}, Paul A. and {Purger}, Norbert and {Quinn}, Thomas and {Raddick}, M. Jordan and {Re Fiorentin}, Paola and {Richards}, Gordon T. and {Richmond}, Michael W. and {Riess}, Adam G. and {Rix}, Hans-Walter and {Rockosi}, Constance M. and {Sako}, Masao and {Schlegel}, David J. and {Schneider}, Donald P. and {Scholz}, Ralf-Dieter and {Schreiber}, Matthias R. and {Schwope}, Axel D. and {Seljak}, Uro{\v{s}} and {Sesar}, Branimir and {Sheldon}, Erin and {Shimasaku}, Kazu and {Sibley}, Valena C. and {Simmons}, A.~E. and {Sivarani}, Thirupathi and {Allyn Smith}, J. and {Smith}, Martin C. and {Smol{\v{c}}i{\'c}}, Vernesa and {Snedden}, Stephanie A. and {Stebbins}, Albert and {Steinmetz}, Matthias and {Stoughton}, Chris and {Strauss}, Michael A. and {SubbaRao}, Mark and {Suto}, Yasushi and {Szalay}, Alexander S. and {Szapudi}, Istv{\'a}n and {Szkody}, Paula and {Tanaka}, Masayuki and {Tegmark}, Max and {Teodoro}, Luis F.~A. and {Thakar}, Aniruddha R. and {Tremonti}, Christy A. and {Tucker}, Douglas L. and {Uomoto}, Alan and {Vanden Berk}, Daniel E. and {Vandenberg}, Jan and {Vidrih}, S. and {Vogeley}, Michael S. and {Voges}, Wolfgang and {Vogt}, Nicole P. and {Wadadekar}, Yogesh and {Watters}, Shannon and {Weinberg}, David H. and {West}, Andrew A. and {White}, Simon D.~M. and {Wilhite}, Brian C. and {Wonders}, Alainna C. and {Yanny}, Brian and {Yocum}, D.~R. and {York}, Donald G. and {Zehavi}, Idit and {Zibetti}, Stefano and {Zucker}, Daniel B.},
        title = "{The Seventh Data Release of the Sloan Digital Sky Survey}",
      journal = {\apjs},
         year = 2009,
        month = jun,
       volume = {182},
       number = {2},
        pages = {543-558},
          doi = {10.1088/0067-0049/182/2/543},
archivePrefix = {arXiv},
       eprint = {0812.0649},
 primaryClass = {astro-ph},
       adsurl = {https://ui.adsabs.harvard.edu/abs/2009ApJS..182..543A}
}

@ARTICLE{nakajima+2022,
       author = {{Nakajima}, Kimihiko and {Ouchi}, Masami and {Xu}, Yi and {Rauch}, Michael and {Harikane}, Yuichi and {Nishigaki}, Moka and {Isobe}, Yuki and {Kusakabe}, Haruka and {Nagao}, Tohru and {Ono}, Yoshiaki and {Onodera}, Masato and {Sugahara}, Yuma and {Kim}, Ji Hoon and {Komiyama}, Yutaka and {Lee}, Chien-Hsiu and {Zahedy}, Fakhri S.},
        title = "{EMPRESS. V. Metallicity Diagnostics of Galaxies over 12 + log(O/H) ≃ 6.9-8.9 Established by a Local Galaxy Census: Preparing for JWST Spectroscopy}",
      journal = {\apjs},
         year = 2022,
        month = sep,
       volume = {262},
       number = {1},
          eid = {3},
        pages = {3},
          doi = {10.3847/1538-4365/ac7710},
archivePrefix = {arXiv},
       eprint = {2206.02824},
 primaryClass = {astro-ph.GA},
       adsurl = {https://ui.adsabs.harvard.edu/abs/2022ApJS..262....3N}
}

@ARTICLE{sanders+2020,
       author = {{Sanders}, Ryan L. and {Shapley}, Alice E. and {Reddy}, Naveen A. and {Kriek}, Mariska and {Siana}, Brian and {Coil}, Alison L. and {Mobasher}, Bahram and {Shivaei}, Irene and {Freeman}, William R. and {Azadi}, Mojegan and {Price}, Sedona H. and {Leung}, Gene and {Fetherolf}, Tara and {de Groot}, Laura and {Zick}, Tom and {Fornasini}, Francesca M. and {Barro}, Guillermo},
        title = "{The MOSDEF survey: direct-method metallicities and ISM conditions at z {\ensuremath{\sim}} 1.5-3.5}",
      journal = {\mnras},
         year = 2020,
        month = jan,
       volume = {491},
       number = {1},
        pages = {1427-1455},
          doi = {10.1093/mnras/stz3032},
archivePrefix = {arXiv},
       eprint = {1907.00013},
 primaryClass = {astro-ph.GA},
       adsurl = {https://ui.adsabs.harvard.edu/abs/2020MNRAS.491.1427S}
}

@ARTICLE{revalski+2024,
       author = {{Revalski}, Mitchell and {Rafelski}, Marc and {Henry}, Alaina and {Fossati}, Matteo and {Fumagalli}, Michele and {Dutta}, Rajeshwari and {Pirzkal}, Norbert and {Beckett}, Alexander and {Arrigoni Battaia}, Fabrizio and {Dayal}, Pratika and {D'Odorico}, Valentina and {Lusso}, Elisabeta and {Nedkova}, Kalina V. and {Prichard}, Laura J. and {Papovich}, Casey and {Peroux}, Celine},
        title = "{The MUSE Ultra Deep Field (MUDF). V. Characterizing the Mass{\textendash}Metallicity Relation for Low-mass Galaxies at z {\ensuremath{\sim}} 1{\textendash}2}",
      journal = {\apj},
         year = 2024,
        month = may,
       volume = {966},
       number = {2},
          eid = {228},
        pages = {228},
          doi = {10.3847/1538-4357/ad382c},
archivePrefix = {arXiv},
       eprint = {2403.17047},
 primaryClass = {astro-ph.GA},
       adsurl = {https://ui.adsabs.harvard.edu/abs/2024ApJ...966..228R}
}

@ARTICLE{curti+2017,
       author = {{Curti}, M. and {Cresci}, G. and {Mannucci}, F. and {Marconi}, A. and {Maiolino}, R. and {Esposito}, S.},
        title = "{New fully empirical calibrations of strong-line metallicity indicators in star-forming galaxies}",
      journal = {\mnras},
         year = 2017,
        month = feb,
       volume = {465},
       number = {2},
        pages = {1384-1400},
          doi = {10.1093/mnras/stw2766},
archivePrefix = {arXiv},
       eprint = {1610.06939},
 primaryClass = {astro-ph.GA},
       adsurl = {https://ui.adsabs.harvard.edu/abs/2017MNRAS.465.1384C}
}

@ARTICLE{galSpecLine1,
       author = {{Tremonti}, Christy A. and {Heckman}, Timothy M. and {Kauffmann}, Guinevere and {Brinchmann}, Jarle and {Charlot}, St{\'e}phane and {White}, Simon D.~M. and {Seibert}, Mark and {Peng}, Eric W. and {Schlegel}, David J. and {Uomoto}, Alan and {Fukugita}, Masataka and {Brinkmann}, Jon},
        title = "{The Origin of the Mass-Metallicity Relation: Insights from 53,000 Star-forming Galaxies in the Sloan Digital Sky Survey}",
      journal = {\apj},
         year = 2004,
        month = oct,
       volume = {613},
       number = {2},
        pages = {898-913},
          doi = {10.1086/423264},
archivePrefix = {arXiv},
       eprint = {astro-ph/0405537},
 primaryClass = {astro-ph},
       adsurl = {https://ui.adsabs.harvard.edu/abs/2004ApJ...613..898T}
}

@ARTICLE{galSpecLine2,
       author = {{Brinchmann}, J. and {Charlot}, S. and {White}, S.~D.~M. and {Tremonti}, C. and {Kauffmann}, G. and {Heckman}, T. and {Brinkmann}, J.},
        title = "{The physical properties of star-forming galaxies in the low-redshift Universe}",
      journal = {\mnras},
         year = 2004,
        month = jul,
       volume = {351},
       number = {4},
        pages = {1151-1179},
          doi = {10.1111/j.1365-2966.2004.07881.x},
archivePrefix = {arXiv},
       eprint = {astro-ph/0311060},
 primaryClass = {astro-ph},
       adsurl = {https://ui.adsabs.harvard.edu/abs/2004MNRAS.351.1151B}
}

@ARTICLE{calzetti+2000,
       author = {{Calzetti}, Daniela and {Armus}, Lee and {Bohlin}, Ralph C. and {Kinney}, Anne L. and {Koornneef}, Jan and {Storchi-Bergmann}, Thaisa},
        title = "{The Dust Content and Opacity of Actively Star-forming Galaxies}",
      journal = {\apj},
         year = 2000,
        month = apr,
       volume = {533},
       number = {2},
        pages = {682-695},
          doi = {10.1086/308692},
archivePrefix = {arXiv},
       eprint = {astro-ph/9911459},
 primaryClass = {astro-ph},
       adsurl = {https://ui.adsabs.harvard.edu/abs/2000ApJ...533..682C}
}

@ARTICLE{1992AJ....103.1330G,
       author = {{Garnett}, Donald R.},
        title = "{Electron Temperature Variations and the Measurement of Nebular Abundances}",
      journal = {\aj},
         year = 1992,
        month = apr,
       volume = {103},
        pages = {1330},
          doi = {10.1086/116146},
       adsurl = {https://ui.adsabs.harvard.edu/abs/1992AJ....103.1330G}
}

@ARTICLE{1982A&AS...48..299S,
       author = {{Stasi{\'n}ska}, G.},
        title = "{A catalogue of model HII regions.}",
      journal = {\aaps},
         year = 1982,
        month = may,
       volume = {48},
        pages = {299-304},
       adsurl = {https://ui.adsabs.harvard.edu/abs/1982A&AS...48..299S}
}

@ARTICLE{2013ApJ...765..140A,
       author = {{Andrews}, Brett H. and {Martini}, Paul},
        title = "{The Mass-Metallicity Relation with the Direct Method on Stacked Spectra of SDSS Galaxies}",
      journal = {\apj},
         year = 2013,
        month = mar,
       volume = {765},
       number = {2},
          eid = {140},
        pages = {140},
          doi = {10.1088/0004-637X/765/2/140},
archivePrefix = {arXiv},
       eprint = {1211.3418},
 primaryClass = {astro-ph.CO},
       adsurl = {https://ui.adsabs.harvard.edu/abs/2013ApJ...765..140A}
}

@ARTICLE{pyneb2,
       author = {{Luridiana}, V. and {Morisset}, C. and {Shaw}, R.~A.},
        title = "{PyNeb: a new tool for analyzing emission lines. I. Code description and validation of results}",
      journal = {\aap},
         year = 2015,
        month = jan,
       volume = {573},
          eid = {A42},
        pages = {A42},
          doi = {10.1051/0004-6361/201323152},
archivePrefix = {arXiv},
       eprint = {1410.6662},
 primaryClass = {astro-ph.IM},
       adsurl = {https://ui.adsabs.harvard.edu/abs/2015A&A...573A..42L}
}

@INPROCEEDINGS{pyneb1,
       author = {{Luridiana}, Valentina and {Morisset}, Christophe and {Shaw}, Richard A.},
        title = "{PyNeb: a new software for the analysis of emission lines}",
    booktitle = {Planetary Nebulae: An Eye to the Future},
         year = 2012,
       series = {IAU Symposium},
       volume = {283},
        month = aug,
        pages = {422-423},
          doi = {10.1017/S1743921312011738},
       adsurl = {https://ui.adsabs.harvard.edu/abs/2012IAUS..283..422L}
}

@ARTICLE{izotov+2006,
       author = {{Izotov}, Y.~I. and {Stasi{\'n}ska}, G. and {Meynet}, G. and {Guseva}, N.~G. and {Thuan}, T.~X.},
        title = "{The chemical composition of metal-poor emission-line galaxies in the Data Release 3 of the Sloan Digital Sky Survey}",
      journal = {\aap},
         year = 2006,
        month = mar,
       volume = {448},
       number = {3},
        pages = {955-970},
          doi = {10.1051/0004-6361:20053763},
archivePrefix = {arXiv},
       eprint = {astro-ph/0511644},
 primaryClass = {astro-ph},
       adsurl = {https://ui.adsabs.harvard.edu/abs/2006A&A...448..955I}
}

@ARTICLE{1986MNRAS.223..811C,
       author = {{Campbell}, Alison and {Terlevich}, Roberto and {Melnick}, Jorge},
        title = "{The stellar populations and evolution of H II galaxies - I. High signal-to-noise optical spectroscopy.}",
      journal = {\mnras},
         year = 1986,
        month = dec,
       volume = {223},
        pages = {811-825},
          doi = {10.1093/mnras/223.4.811},
       adsurl = {https://ui.adsabs.harvard.edu/abs/1986MNRAS.223..811C}
}

@ARTICLE{2006MNRAS.370.1928P,
       author = {{Pilyugin}, Leonid S. and {V{\'\i}lchez}, Jos{\'e} M. and {Thuan}, Trinh X.},
        title = "{On the relation between electron temperatures in the O$^{+}$ and O$^{++}$ zones in high-metallicity HII regions}",
      journal = {\mnras},
         year = 2006,
        month = aug,
       volume = {370},
       number = {4},
        pages = {1928-1934},
          doi = {10.1111/j.1365-2966.2006.10618.x},
archivePrefix = {arXiv},
       eprint = {astro-ph/0605695},
 primaryClass = {astro-ph},
       adsurl = {https://ui.adsabs.harvard.edu/abs/2006MNRAS.370.1928P}
}

@ARTICLE{2006MNRAS.367.1139P,
       author = {{Pilyugin}, Leonid S. and {Thuan}, Trinh X. and {V{\'\i}lchez}, Jos{\'e} M.},
        title = "{Oxygen abundances in the most oxygen-rich spiral galaxies}",
      journal = {\mnras},
         year = 2006,
        month = apr,
       volume = {367},
       number = {3},
        pages = {1139-1146},
          doi = {10.1111/j.1365-2966.2006.10033.x},
archivePrefix = {arXiv},
       eprint = {astro-ph/0601122},
 primaryClass = {astro-ph},
       adsurl = {https://ui.adsabs.harvard.edu/abs/2006MNRAS.367.1139P}
}

@ARTICLE{2009MNRAS.398..485P,
       author = {{Pilyugin}, L.~S. and {Mattsson}, L. and {V{\'\i}lchez}, J.~M. and {Cedr{\'e}s}, B.},
        title = "{On the electron temperatures in high-metallicity HII regions}",
      journal = {\mnras},
         year = 2009,
        month = sep,
       volume = {398},
       number = {1},
        pages = {485-496},
          doi = {10.1111/j.1365-2966.2009.15182.x},
archivePrefix = {arXiv},
       eprint = {0907.0084},
 primaryClass = {astro-ph.CO},
       adsurl = {https://ui.adsabs.harvard.edu/abs/2009MNRAS.398..485P}
}

@ARTICLE{2010ApJ...720.1738P,
       author = {{Pilyugin}, Leonid S. and {V{\'\i}lchez}, Jos{\'e} M. and {Thuan}, Trinh X.},
        title = "{New Improved Calibration Relations for the Determination of Electron Temperatures and Oxygen and Nitrogen Abundances in H II Regions}",
      journal = {\apj},
         year = 2010,
        month = sep,
       volume = {720},
       number = {2},
        pages = {1738-1751},
          doi = {10.1088/0004-637X/720/2/1738},
       adsurl = {https://ui.adsabs.harvard.edu/abs/2010ApJ...720.1738P}
}

@BOOK{1992mde..book.....S,
       author = {{Scott}, D.~W.},
        title = "{Multivariate Density Estimation}",
         year = 1992,
       adsurl = {https://ui.adsabs.harvard.edu/abs/1992mde..book.....S}
}

@BOOK{1986desd.book.....S,
       author = {{Silverman}, B.~W.},
        title = "{Density estimation for statistics and data analysis}",
         year = 1986,
       adsurl = {https://ui.adsabs.harvard.edu/abs/1986desd.book.....S}
}

@ARTICLE{scipy,
       author = {{Virtanen}, Pauli and {Gommers}, Ralf and {Oliphant}, Travis E. and {Haberland}, Matt and {Reddy}, Tyler and {Cournapeau}, David and {Burovski}, Evgeni and {Peterson}, Pearu and {Weckesser}, Warren and {Bright}, Jonathan and {van der Walt}, St{\'e}fan J. and {Brett}, Matthew and {Wilson}, Joshua and {Millman}, K. Jarrod and {Mayorov}, Nikolay and {Nelson}, Andrew R.~J. and {Jones}, Eric and {Kern}, Robert and {Larson}, Eric and {Carey}, C.~J. and {Polat}, {\.I}lhan and {Feng}, Yu and {Moore}, Eric W. and {VanderPlas}, Jake and {Laxalde}, Denis and {Perktold}, Josef and {Cimrman}, Robert and {Henriksen}, Ian and {Quintero}, E.~A. and {Harris}, Charles R. and {Archibald}, Anne M. and {Ribeiro}, Ant{\^o}nio H. and {Pedregosa}, Fabian and {van Mulbregt}, Paul and {SciPy 1. 0 Contributors}},
        title = "{SciPy 1.0: fundamental algorithms for scientific computing in Python}",
      journal = {Nature Methods},
         year = 2020,
        month = feb,
       volume = {17},
        pages = {261-272},
          doi = {10.1038/s41592-019-0686-2},
archivePrefix = {arXiv},
       eprint = {1907.10121},
 primaryClass = {cs.MS},
       adsurl = {https://ui.adsabs.harvard.edu/abs/2020NatMe..17..261V}
}

@ARTICLE{isobe+2023,
       author = {{Isobe}, Yuki and {Ouchi}, Masami and {Nakajima}, Kimihiko and {Harikane}, Yuichi and {Ono}, Yoshiaki and {Xu}, Yi and {Zhang}, Yechi and {Umeda}, Hiroya},
        title = "{Redshift Evolution of Electron Density in the Interstellar Medium at z   0-9 Uncovered with JWST/NIRSpec Spectra and Line-spread Function Determinations}",
      journal = {\apj},
         year = 2023,
        month = oct,
       volume = {956},
       number = {2},
          eid = {139},
        pages = {139},
          doi = {10.3847/1538-4357/acf376},
archivePrefix = {arXiv},
       eprint = {2301.06811},
 primaryClass = {astro-ph.GA},
       adsurl = {https://ui.adsabs.harvard.edu/abs/2023ApJ...956..139I}
}

@ARTICLE{2023MNRAS.526.3504H,
       author = {{Hirschmann}, Michaela and {Charlot}, Stephane and {Somerville}, Rachel S.},
        title = "{High-redshift metallicity calibrations for JWST spectra: insights from line emission in cosmological simulations}",
      journal = {\mnras},
         year = 2023,
        month = dec,
       volume = {526},
       number = {3},
        pages = {3504-3518},
          doi = {10.1093/mnras/stad2745},
archivePrefix = {arXiv},
       eprint = {2305.03753},
 primaryClass = {astro-ph.GA},
       adsurl = {https://ui.adsabs.harvard.edu/abs/2023MNRAS.526.3504H}
}

@ARTICLE{2002ApJS..142...35K,
       author = {{Kewley}, L.~J. and {Dopita}, M.~A.},
        title = "{Using Strong Lines to Estimate Abundances in Extragalactic H II Regions and Starburst Galaxies}",
      journal = {\apjs},
         year = 2002,
        month = sep,
       volume = {142},
       number = {1},
        pages = {35-52},
          doi = {10.1086/341326},
archivePrefix = {arXiv},
       eprint = {astro-ph/0206495},
 primaryClass = {astro-ph},
       adsurl = {https://ui.adsabs.harvard.edu/abs/2002ApJS..142...35K}
}

@ARTICLE{2018A&A...612A..94N,
       author = {{Nakajima}, K. and {Schaerer}, D. and {Le F{\`e}vre}, O. and {Amor{\'\i}n}, R. and {Talia}, M. and {Lemaux}, B.~C. and {Tasca}, L.~A.~M. and {Vanzella}, E. and {Zamorani}, G. and {Bardelli}, S. and {Grazian}, A. and {Guaita}, L. and {Hathi}, N.~P. and {Pentericci}, L. and {Zucca}, E.},
        title = "{The VIMOS Ultra Deep Survey: Nature, ISM properties, and ionizing spectra of CIII]{\ensuremath{\lambda}}1909 emitters at z = 2-4}",
      journal = {\aap},
         year = 2018,
        month = may,
       volume = {612},
          eid = {A94},
        pages = {A94},
          doi = {10.1051/0004-6361/201731935},
archivePrefix = {arXiv},
       eprint = {1709.03990},
 primaryClass = {astro-ph.GA},
       adsurl = {https://ui.adsabs.harvard.edu/abs/2018A&A...612A..94N}
}

@ARTICLE{2023NatAs...7.1517H,
       author = {{Heintz}, Kasper E. and {Brammer}, Gabriel B. and {Gim{\'e}nez-Arteaga}, Clara and {Strait}, Victoria B. and {del P. Lagos}, Claudia and {Vijayan}, Aswin P. and {Matthee}, Jorryt and {Watson}, Darach and {Mason}, Charlotte A. and {Hutter}, Anne and {Toft}, Sune and {Fynbo}, Johan P.~U. and {Oesch}, Pascal A.},
        title = "{Dilution of chemical enrichment in galaxies 600 Myr after the Big Bang}",
      journal = {Nature Astronomy},
         year = 2023,
        month = dec,
       volume = {7},
        pages = {1517-1524},
          doi = {10.1038/s41550-023-02078-7},
archivePrefix = {arXiv},
       eprint = {2212.02890},
 primaryClass = {astro-ph.GA},
       adsurl = {https://ui.adsabs.harvard.edu/abs/2023NatAs...7.1517H}
}

@ARTICLE{EVOLFMR,
       author = {{Langeroodi}, Danial and {Hjorth}, Jens},
        title = "{Ultraviolet Compactness of High-Redshift Galaxies as a Tracer of Early-Stage Gas Infall, Stochastic Star Formation, and Offset from the Fundamental Metallicity Relation}",
      journal = {arXiv e-prints},
         year = 2023,
        month = jul,
          eid = {arXiv:2307.06336},
        pages = {arXiv:2307.06336},
          doi = {10.48550/arXiv.2307.06336},
archivePrefix = {arXiv},
       eprint = {2307.06336},
 primaryClass = {astro-ph.GA},
       adsurl = {https://ui.adsabs.harvard.edu/abs/2023arXiv230706336L}
}

@ARTICLE{2024arXiv240717110C,
       author = {{Chemerynska}, Iryna and {Atek}, Hakim and {Dayal}, Pratika and {Furtak}, Lukas J. and {Feldmann}, Robert and {Greene}, Jenny E. and {Maseda}, Michael V. and {Nanayakkara}, Themiya and {Oesch}, Pascal A. and {Labbe}, Ivo and {Bezanson}, Rachel and {Brammer}, Gabriel and {Cutler}, Sam E. and {Leja}, Joel and {Pan}, Richard and {Price}, Sedona H. and {Wang}, Bingjie and {Weaver}, John R. and {Whitaker}, Katherine E.},
        title = "{The Extreme Low-mass End of the Mass-Metallicity Relation at $z\sim7$}",
      journal = {arXiv e-prints},
         year = 2024,
        month = jul,
          eid = {arXiv:2407.17110},
        pages = {arXiv:2407.17110},
          doi = {10.48550/arXiv.2407.17110},
archivePrefix = {arXiv},
       eprint = {2407.17110},
 primaryClass = {astro-ph.GA},
       adsurl = {https://ui.adsabs.harvard.edu/abs/2024arXiv240717110C}
}

@ARTICLE{2024arXiv240807974S,
       author = {{Sarkar}, Arnab and {Chakraborty}, Priyanka and {Vogelsberger}, Mark and {McDonald}, Michael and {Torrey}, Paul and {Garcia}, Alex M. and {Khullar}, Gourav and {Ferland}, Gary J. and {Forman}, William and {Wolk}, Scott and {Schneider}, Benjamin and {Bautz}, Mark and {Miller}, Eric and {Grant}, Catherine and {ZuHone}, John},
        title = "{Unveiling the Cosmic Chemistry: Revisiting the Mass-Metallicity Relation with JWST/NIRSpec at 4 < z < 10}",
      journal = {arXiv e-prints},
         year = 2024,
        month = aug,
          eid = {arXiv:2408.07974},
        pages = {arXiv:2408.07974},
          doi = {10.48550/arXiv.2408.07974},
archivePrefix = {arXiv},
       eprint = {2408.07974},
 primaryClass = {astro-ph.GA},
       adsurl = {https://ui.adsabs.harvard.edu/abs/2024arXiv240807974S}
}

@ARTICLE{ZEIGHT,
       author = {{Langeroodi}, Danial and {Hjorth}, Jens and {Chen}, Wenlei and {Kelly}, Patrick L. and {Williams}, Hayley and {Lin}, Yu-Heng and {Scarlata}, Claudia and {Zitrin}, Adi and {Broadhurst}, Tom and {Diego}, Jose M. and {Huang}, Xiaosheng and {Filippenko}, Alexei V. and {Foley}, Ryan J. and {Jha}, Saurabh and {Koekemoer}, Anton M. and {Oguri}, Masamune and {Perez-Fournon}, Ismael and {Pierel}, Justin and {Poidevin}, Frederick and {Strolger}, Lou},
        title = "{Evolution of the Mass-Metallicity Relation from Redshift z {\ensuremath{\approx}} 8 to the Local Universe}",
      journal = {\apj},
         year = 2023,
        month = nov,
       volume = {957},
       number = {1},
          eid = {39},
        pages = {39},
          doi = {10.3847/1538-4357/acdbc1},
archivePrefix = {arXiv},
       eprint = {2212.02491},
 primaryClass = {astro-ph.GA},
       adsurl = {https://ui.adsabs.harvard.edu/abs/2023ApJ...957...39L}
}

@ARTICLE{2023ApJ...952..143R,
       author = {{Rinaldi}, P. and {Caputi}, K.~I. and {Costantin}, L. and {Gillman}, S. and {Iani}, E. and {P{\'e}rez-Gonz{\'a}lez}, P.~G. and {{\"O}stlin}, G. and {Colina}, L. and {Greve}, T.~R. and {Noorgard-Nielsen}, H.~U. and {Wright}, G.~S. and {Alonso-Herrero}, A. and {{\'A}lvarez-M{\'a}rquez}, J. and {Eckart}, A. and {Garc{\'\i}a-Mar{\'\i}n}, M. and {Hjorth}, J. and {Ilbert}, O. and {Kendrew}, S. and {Labiano}, A. and {Le F{\`e}vre}, O. and {Pye}, J. and {Tikkanen}, T. and {Walter}, F. and {van der Werf}, P. and {Ward}, M. and {Annunziatella}, M. and {Azzollini}, R. and {Bik}, A. and {Boogaard}, L. and {Bosman}, S.~E.~I. and {Crespo G{\'o}mez}, A. and {Jermann}, I. and {Langeroodi}, D. and {Melinder}, J. and {Meyer}, R.~A. and {Moutard}, T. and {Peissker}, F. and {Topinka}, M. and {van Dishoeck}, E. and {G{\"u}del}, M. and {Henning}, Th. and {Lagage}, P. -O. and {Ray}, T. and {Vandenbussche}, B. and {Waelkens}, C. and {Navarro-Carrera}, R. and {Kokorev}, V.},
        title = "{MIDIS: Strong (H{\ensuremath{\beta}}+[O III]) and H{\ensuremath{\alpha}} Emitters at Redshift z ≃ 7-8 Unveiled with JWST NIRCam and MIRI Imaging in the Hubble eXtreme Deep Field}",
      journal = {\apj},
         year = 2023,
        month = aug,
       volume = {952},
       number = {2},
          eid = {143},
        pages = {143},
          doi = {10.3847/1538-4357/acdc27},
archivePrefix = {arXiv},
       eprint = {2301.10717},
 primaryClass = {astro-ph.GA},
       adsurl = {https://ui.adsabs.harvard.edu/abs/2023ApJ...952..143R}
}

@ARTICLE{2016ApJ...833..254S,
       author = {{Smit}, Renske and {Bouwens}, Rychard J. and {Labb{\'e}}, Ivo and {Franx}, Marijn and {Wilkins}, Stephen M. and {Oesch}, Pascal A.},
        title = "{Inferred H⍺ Flux as a Star Formation Rate Indicator at z \raisebox{-0.5ex}\textasciitilde 4-5: Implications for Dust Properties, Burstiness, and the z = 4-8 Star Formation Rate Functions}",
      journal = {\apj},
         year = 2016,
        month = dec,
       volume = {833},
       number = {2},
          eid = {254},
        pages = {254},
          doi = {10.3847/1538-4357/833/2/254},
archivePrefix = {arXiv},
       eprint = {1511.08808},
 primaryClass = {astro-ph.GA},
       adsurl = {https://ui.adsabs.harvard.edu/abs/2016ApJ...833..254S}
}

@ARTICLE{jakobsen+2022,
       author = {{Jakobsen}, P. and {Ferruit}, P. and {Alves de Oliveira}, C. and {Arribas}, S. and {Bagnasco}, G. and {Barho}, R. and {Beck}, T.~L. and {Birkmann}, S. and {B{\"o}ker}, T. and {Bunker}, A.~J. and {Charlot}, S. and {de Jong}, P. and {de Marchi}, G. and {Ehrenwinkler}, R. and {Falcolini}, M. and {Fels}, R. and {Franx}, M. and {Franz}, D. and {Funke}, M. and {Giardino}, G. and {Gnata}, X. and {Holota}, W. and {Honnen}, K. and {Jensen}, P.~L. and {Jentsch}, M. and {Johnson}, T. and {Jollet}, D. and {Karl}, H. and {Kling}, G. and {K{\"o}hler}, J. and {Kolm}, M. -G. and {Kumari}, N. and {Lander}, M.~E. and {Lemke}, R. and {L{\'o}pez-Caniego}, M. and {L{\"u}tzgendorf}, N. and {Maiolino}, R. and {Manjavacas}, E. and {Marston}, A. and {Maschmann}, M. and {Maurer}, R. and {Messerschmidt}, B. and {Moseley}, S.~H. and {Mosner}, P. and {Mott}, D.~B. and {Muzerolle}, J. and {Pirzkal}, N. and {Pittet}, J. -F. and {Plitzke}, A. and {Posselt}, W. and {Rapp}, B. and {Rauscher}, B.~J. and {Rawle}, T. and {Rix}, H. -W. and {R{\"o}del}, A. and {Rumler}, P. and {Sabbi}, E. and {Salvignol}, J. -C. and {Schmid}, T. and {Sirianni}, M. and {Smith}, C. and {Strada}, P. and {te Plate}, M. and {Valenti}, J. and {Wettemann}, T. and {Wiehe}, T. and {Wiesmayer}, M. and {Willott}, C.~J. and {Wright}, R. and {Zeidler}, P. and {Zincke}, C.},
        title = "{The Near-Infrared Spectrograph (NIRSpec) on the James Webb Space Telescope. I. Overview of the instrument and its capabilities}",
      journal = {\aap},
         year = 2022,
        month = may,
       volume = {661},
          eid = {A80},
        pages = {A80},
          doi = {10.1051/0004-6361/202142663},
archivePrefix = {arXiv},
       eprint = {2202.03305},
 primaryClass = {astro-ph.IM},
       adsurl = {https://ui.adsabs.harvard.edu/abs/2022A&A...661A..80J}
}

@ARTICLE{ferruit+2022,
       author = {{Ferruit}, P. and {Jakobsen}, P. and {Giardino}, G. and {Rawle}, T. and {Alves de Oliveira}, C. and {Arribas}, S. and {Beck}, T.~L. and {Birkmann}, S. and {B{\"o}ker}, T. and {Bunker}, A.~J. and {Charlot}, S. and {de Marchi}, G. and {Franx}, M. and {Henry}, A. and {Karakla}, D. and {Kassin}, S.~A. and {Kumari}, N. and {L{\'o}pez-Caniego}, M. and {L{\"u}tzgendorf}, N. and {Maiolino}, R. and {Manjavacas}, E. and {Marston}, A. and {Moseley}, S.~H. and {Muzerolle}, J. and {Pirzkal}, N. and {Rauscher}, B. and {Rix}, H. -W. and {Sabbi}, E. and {Sirianni}, M. and {te Plate}, M. and {Valenti}, J. and {Willott}, C.~J. and {Zeidler}, P.},
        title = "{The Near-Infrared Spectrograph (NIRSpec) on the James Webb Space Telescope. II. Multi-object spectroscopy (MOS)}",
      journal = {\aap},
         year = 2022,
        month = may,
       volume = {661},
          eid = {A81},
        pages = {A81},
          doi = {10.1051/0004-6361/202142673},
archivePrefix = {arXiv},
       eprint = {2202.03306},
 primaryClass = {astro-ph.IM},
       adsurl = {https://ui.adsabs.harvard.edu/abs/2022A&A...661A..81F}
}

@ARTICLE{sanders+2015,
       author = {{Sanders}, Ryan L. and {Shapley}, Alice E. and {Kriek}, Mariska and {Reddy}, Naveen A. and {Freeman}, William R. and {Coil}, Alison L. and {Siana}, Brian and {Mobasher}, Bahram and {Shivaei}, Irene and {Price}, Sedona H. and {de Groot}, Laura},
        title = "{The MOSDEF Survey: Mass, Metallicity, and Star-formation Rate at z \raisebox{-0.5ex}\textasciitilde 2.3}",
      journal = {\apj},
         year = 2015,
        month = feb,
       volume = {799},
       number = {2},
          eid = {138},
        pages = {138},
          doi = {10.1088/0004-637X/799/2/138},
archivePrefix = {arXiv},
       eprint = {1408.2521},
 primaryClass = {astro-ph.GA},
       adsurl = {https://ui.adsabs.harvard.edu/abs/2015ApJ...799..138S}
}

@ARTICLE{sanders+2021,
       author = {{Sanders}, Ryan L. and {Shapley}, Alice E. and {Jones}, Tucker and {Reddy}, Naveen A. and {Kriek}, Mariska and {Siana}, Brian and {Coil}, Alison L. and {Mobasher}, Bahram and {Shivaei}, Irene and {Dav{\'e}}, Romeel and {Azadi}, Mojegan and {Price}, Sedona H. and {Leung}, Gene and {Freeman}, William R. and {Fetherolf}, Tara and {de Groot}, Laura and {Zick}, Tom and {Barro}, Guillermo},
        title = "{The MOSDEF Survey: The Evolution of the Mass-Metallicity Relation from z = 0 to z 3.3}",
      journal = {\apj},
         year = 2021,
        month = jun,
       volume = {914},
       number = {1},
          eid = {19},
        pages = {19},
          doi = {10.3847/1538-4357/abf4c1},
archivePrefix = {arXiv},
       eprint = {2009.07292},
 primaryClass = {astro-ph.GA},
       adsurl = {https://ui.adsabs.harvard.edu/abs/2021ApJ...914...19S}
}

@ARTICLE{curti+2024,
       author = {{Curti}, Mirko and {Maiolino}, Roberto and {Curtis-Lake}, Emma and {Chevallard}, Jacopo and {Carniani}, Stefano and {D'Eugenio}, Francesco and {Looser}, Tobias J. and {Scholtz}, Jan and {Charlot}, Stephane and {Cameron}, Alex and {{\"U}bler}, Hannah and {Witstok}, Joris and {Boyett}, Kristian and {Laseter}, Isaac and {Sandles}, Lester and {Arribas}, Santiago and {Bunker}, Andrew and {Giardino}, Giovanna and {Maseda}, Michael V. and {Rawle}, Tim and {Rodr{\'\i}guez Del Pino}, Bruno and {Smit}, Renske and {Willott}, Chris J. and {Eisenstein}, Daniel J. and {Hausen}, Ryan and {Johnson}, Benjamin and {Rieke}, Marcia and {Robertson}, Brant and {Tacchella}, Sandro and {Williams}, Christina C. and {Willmer}, Christopher and {Baker}, William M. and {Bhatawdekar}, Rachana and {Egami}, Eiichi and {Helton}, Jakob M. and {Ji}, Zhiyuan and {Kumari}, Nimisha and {Perna}, Michele and {Shivaei}, Irene and {Sun}, Fengwu},
        title = "{JADES: Insights into the low-mass end of the mass-metallicity-SFR relation at 3 < z < 10 from deep JWST/NIRSpec spectroscopy}",
      journal = {\aap},
         year = 2024,
        month = apr,
       volume = {684},
          eid = {A75},
        pages = {A75},
          doi = {10.1051/0004-6361/202346698},
archivePrefix = {arXiv},
       eprint = {2304.08516},
 primaryClass = {astro-ph.GA},
       adsurl = {https://ui.adsabs.harvard.edu/abs/2024A&A...684A..75C}
}

@ARTICLE{2005ApJ...635..260S,
       author = {{Savaglio}, S. and {Glazebrook}, K. and {Le Borgne}, D. and {Juneau}, S. and {Abraham}, R.~G. and {Chen}, H. -W. and {Crampton}, D. and {McCarthy}, P.~J. and {Carlberg}, R.~G. and {Marzke}, R.~O. and {Roth}, K. and {J{\o}rgensen}, I. and {Murowinski}, R.},
        title = "{The Gemini Deep Deep Survey. VII. The Redshift Evolution of the Mass-Metallicity Relation}",
      journal = {\apj},
         year = 2005,
        month = dec,
       volume = {635},
       number = {1},
        pages = {260-279},
          doi = {10.1086/497331},
archivePrefix = {arXiv},
       eprint = {astro-ph/0508407},
 primaryClass = {astro-ph},
       adsurl = {https://ui.adsabs.harvard.edu/abs/2005ApJ...635..260S}
}

@ARTICLE{2006ApJ...644..813E,
       author = {{Erb}, Dawn K. and {Shapley}, Alice E. and {Pettini}, Max and {Steidel}, Charles C. and {Reddy}, Naveen A. and {Adelberger}, Kurt L.},
        title = "{The Mass-Metallicity Relation at z>\raisebox{-0.5ex}\textasciitilde2}",
      journal = {\apj},
         year = 2006,
        month = jun,
       volume = {644},
       number = {2},
        pages = {813-828},
          doi = {10.1086/503623},
archivePrefix = {arXiv},
       eprint = {astro-ph/0602473},
 primaryClass = {astro-ph},
       adsurl = {https://ui.adsabs.harvard.edu/abs/2006ApJ...644..813E}
}

@ARTICLE{2008A&A...488..463M,
       author = {{Maiolino}, R. and {Nagao}, T. and {Grazian}, A. and {Cocchia}, F. and {Marconi}, A. and {Mannucci}, F. and {Cimatti}, A. and {Pipino}, A. and {Ballero}, S. and {Calura}, F. and {Chiappini}, C. and {Fontana}, A. and {Granato}, G.~L. and {Matteucci}, F. and {Pastorini}, G. and {Pentericci}, L. and {Risaliti}, G. and {Salvati}, M. and {Silva}, L.},
        title = "{AMAZE. I. The evolution of the mass-metallicity relation at z > 3}",
      journal = {\aap},
         year = 2008,
        month = sep,
       volume = {488},
       number = {2},
        pages = {463-479},
          doi = {10.1051/0004-6361:200809678},
archivePrefix = {arXiv},
       eprint = {0806.2410},
 primaryClass = {astro-ph},
       adsurl = {https://ui.adsabs.harvard.edu/abs/2008A&A...488..463M}
}

@ARTICLE{2009MNRAS.398.1915M,
       author = {{Mannucci}, F. and {Cresci}, G. and {Maiolino}, R. and {Marconi}, A. and {Pastorini}, G. and {Pozzetti}, L. and {Gnerucci}, A. and {Risaliti}, G. and {Schneider}, R. and {Lehnert}, M. and {Salvati}, M.},
        title = "{LSD: Lyman-break galaxies Stellar populations and Dynamics - I. Mass, metallicity and gas at z \raisebox{-0.5ex}\textasciitilde 3.1}",
      journal = {\mnras},
         year = 2009,
        month = oct,
       volume = {398},
       number = {4},
        pages = {1915-1931},
          doi = {10.1111/j.1365-2966.2009.15185.x},
archivePrefix = {arXiv},
       eprint = {0902.2398},
 primaryClass = {astro-ph.CO},
       adsurl = {https://ui.adsabs.harvard.edu/abs/2009MNRAS.398.1915M}
}

@ARTICLE{2011ApJ...730..137Z,
       author = {{Zahid}, H.~J. and {Kewley}, L.~J. and {Bresolin}, F.},
        title = "{The Mass-Metallicity and Luminosity-Metallicity Relations from DEEP2 at z \raisebox{-0.5ex}\textasciitilde 0.8}",
      journal = {\apj},
         year = 2011,
        month = apr,
       volume = {730},
       number = {2},
          eid = {137},
        pages = {137},
          doi = {10.1088/0004-637X/730/2/137},
archivePrefix = {arXiv},
       eprint = {1006.4877},
 primaryClass = {astro-ph.CO},
       adsurl = {https://ui.adsabs.harvard.edu/abs/2011ApJ...730..137Z}
}

@ARTICLE{2014ApJ...792...75Z,
       author = {{Zahid}, H.~J. and {Kashino}, D. and {Silverman}, J.~D. and {Kewley}, L.~J. and {Daddi}, E. and {Renzini}, A. and {Rodighiero}, G. and {Nagao}, T. and {Arimoto}, N. and {Sanders}, D.~B. and {Kartaltepe}, J. and {Lilly}, S.~J. and {Maier}, C. and {Geller}, M.~J. and {Capak}, P. and {Carollo}, C.~M. and {Chu}, J. and {Hasinger}, G. and {Ilbert}, O. and {Kajisawa}, M. and {Koekemoer}, A.~M. and {Kovacs}, K. and {Le F{\`e}vre}, O. and {Masters}, D. and {McCracken}, H.~J. and {Onodera}, M. and {Scoville}, N. and {Strazzullo}, V. and {Sugiyama}, N. and {Taniguchi}, Y. and {COSMOS Team}},
        title = "{The FMOS-COSMOS Survey of Star-forming Galaxies at z \raisebox{-0.5ex}\textasciitilde 1.6. II. The Mass-Metallicity Relation and the Dependence on Star Formation Rate and Dust Extinction}",
      journal = {\apj},
         year = 2014,
        month = sep,
       volume = {792},
       number = {1},
          eid = {75},
        pages = {75},
          doi = {10.1088/0004-637X/792/1/75},
archivePrefix = {arXiv},
       eprint = {1310.4950},
 primaryClass = {astro-ph.CO},
       adsurl = {https://ui.adsabs.harvard.edu/abs/2014ApJ...792...75Z}
}

@ARTICLE{2016ApJ...827...74W,
       author = {{Wuyts}, Eva and {Wisnioski}, Emily and {Fossati}, Matteo and {F{\"o}rster Schreiber}, Natascha M. and {Genzel}, Reinhard and {Davies}, Ric and {Mendel}, J. Trevor and {Naab}, Thorsten and {R{\"o}ttgers}, Bernhard and {Wilman}, David J. and {Wuyts}, Stijn and {Bandara}, Kaushala and {Beifiori}, Alessandra and {Belli}, Sirio and {Bender}, Ralf and {Brammer}, Gabriel B. and {Burkert}, Andreas and {Chan}, Jeffrey and {Galametz}, Audrey and {Kulkarni}, Sandesh K. and {Lang}, Philipp and {Lutz}, Dieter and {Momcheva}, Ivelina G. and {Nelson}, Erica J. and {Rosario}, David and {Saglia}, Roberto P. and {Seitz}, Stella and {Tacconi}, Linda J. and {Tadaki}, Ken-ichi and {{\"U}bler}, Hannah and {van Dokkum}, Pieter},
        title = "{The Evolution of Metallicity and Metallicity Gradients from z = 2.7 to 0.6 with KMOS$^{3D}$}",
      journal = {\apj},
         year = 2016,
        month = aug,
       volume = {827},
       number = {1},
          eid = {74},
        pages = {74},
          doi = {10.3847/0004-637X/827/1/74},
archivePrefix = {arXiv},
       eprint = {1603.01139},
 primaryClass = {astro-ph.GA},
       adsurl = {https://ui.adsabs.harvard.edu/abs/2016ApJ...827...74W}
}

@ARTICLE{2012ApJ...755...73W,
       author = {{Wuyts}, Eva and {Rigby}, Jane R. and {Sharon}, Keren and {Gladders}, Michael D.},
        title = "{Constraints on the Low-mass End of the Mass-Metallicity Relation at z = 1-2 from Lensed Galaxies}",
      journal = {\apj},
         year = 2012,
        month = aug,
       volume = {755},
       number = {1},
          eid = {73},
        pages = {73},
          doi = {10.1088/0004-637X/755/1/73},
archivePrefix = {arXiv},
       eprint = {1202.5267},
 primaryClass = {astro-ph.CO},
       adsurl = {https://ui.adsabs.harvard.edu/abs/2012ApJ...755...73W}
}

@ARTICLE{2013ApJ...772..141B,
       author = {{Belli}, Sirio and {Jones}, Tucker and {Ellis}, Richard S. and {Richard}, Johan},
        title = "{Testing the Universality of the Fundamental Metallicity Relation at High Redshift Using Low-mass Gravitationally Lensed Galaxies}",
      journal = {\apj},
         year = 2013,
        month = aug,
       volume = {772},
       number = {2},
          eid = {141},
        pages = {141},
          doi = {10.1088/0004-637X/772/2/141},
archivePrefix = {arXiv},
       eprint = {1302.3614},
 primaryClass = {astro-ph.CO},
       adsurl = {https://ui.adsabs.harvard.edu/abs/2013ApJ...772..141B}
}

@ARTICLE{2013ApJ...776L..27H,
       author = {{Henry}, Alaina and {Scarlata}, Claudia and {Dom{\'\i}nguez}, Alberto and {Malkan}, Matthew and {Martin}, Crystal L. and {Siana}, Brian and {Atek}, Hakim and {Bedregal}, Alejandro G. and {Colbert}, James W. and {Rafelski}, Marc and {Ross}, Nathaniel and {Teplitz}, Harry and {Bunker}, Andrew J. and {Dressler}, Alan and {Hathi}, Nimish and {Masters}, Daniel and {McCarthy}, Patrick and {Straughn}, Amber},
        title = "{Low Masses and High Redshifts: The Evolution of the Mass-Metallicity Relation}",
      journal = {\apjl},
         year = 2013,
        month = oct,
       volume = {776},
       number = {2},
          eid = {L27},
        pages = {L27},
          doi = {10.1088/2041-8205/776/2/L27},
archivePrefix = {arXiv},
       eprint = {1309.4458},
 primaryClass = {astro-ph.CO},
       adsurl = {https://ui.adsabs.harvard.edu/abs/2013ApJ...776L..27H}
}

@ARTICLE{2013ApJ...774..130K,
       author = {{Kulas}, Kristin R. and {McLean}, Ian S. and {Shapley}, Alice E. and {Steidel}, Charles C. and {Konidaris}, Nicholas P. and {Matthews}, Keith and {Mace}, Gregory N. and {Rudie}, Gwen C. and {Trainor}, Ryan F. and {Reddy}, Naveen A.},
        title = "{The Mass-Metallicity Relation of a z \raisebox{-0.5ex}\textasciitilde 2 Protocluster with MOSFIRE}",
      journal = {\apj},
         year = 2013,
        month = sep,
       volume = {774},
       number = {2},
          eid = {130},
        pages = {130},
          doi = {10.1088/0004-637X/774/2/130},
archivePrefix = {arXiv},
       eprint = {1306.6334},
 primaryClass = {astro-ph.CO},
       adsurl = {https://ui.adsabs.harvard.edu/abs/2013ApJ...774..130K}
}

@ARTICLE{2014MNRAS.440.2300C,
       author = {{Cullen}, F. and {Cirasuolo}, M. and {McLure}, R.~J. and {Dunlop}, J.~S. and {Bowler}, R.~A.~A.},
        title = "{The mass-metallicity-star formation rate relation at z {\ensuremath{\gtrsim}} 2 with 3D Hubble Space Telescope}",
      journal = {\mnras},
         year = 2014,
        month = may,
       volume = {440},
       number = {3},
        pages = {2300-2312},
          doi = {10.1093/mnras/stu443},
archivePrefix = {arXiv},
       eprint = {1310.0816},
 primaryClass = {astro-ph.CO},
       adsurl = {https://ui.adsabs.harvard.edu/abs/2014MNRAS.440.2300C}
}

@ARTICLE{2014ApJ...792....3M,
       author = {{Maier}, C. and {Lilly}, S.~J. and {Ziegler}, B.~L. and {Contini}, T. and {P{\'e}rez Montero}, E. and {Peng}, Y. and {Balestra}, I.},
        title = "{The Mass-Metallicity and Fundamental Metallicity Relations at z > 2 Using Very Large Telescope and Subaru Near-infrared Spectroscopy of zCOSMOS Galaxies}",
      journal = {\apj},
         year = 2014,
        month = sep,
       volume = {792},
       number = {1},
          eid = {3},
        pages = {3},
          doi = {10.1088/0004-637X/792/1/3},
archivePrefix = {arXiv},
       eprint = {1406.6069},
 primaryClass = {astro-ph.GA},
       adsurl = {https://ui.adsabs.harvard.edu/abs/2014ApJ...792....3M}
}

@ARTICLE{2014ApJ...795..165S,
       author = {{Steidel}, Charles C. and {Rudie}, Gwen C. and {Strom}, Allison L. and {Pettini}, Max and {Reddy}, Naveen A. and {Shapley}, Alice E. and {Trainor}, Ryan F. and {Erb}, Dawn K. and {Turner}, Monica L. and {Konidaris}, Nicholas P. and {Kulas}, Kristin R. and {Mace}, Gregory and {Matthews}, Keith and {McLean}, Ian S.},
        title = "{Strong Nebular Line Ratios in the Spectra of z \raisebox{-0.5ex}\textasciitilde 2-3 Star Forming Galaxies: First Results from KBSS-MOSFIRE}",
      journal = {\apj},
         year = 2014,
        month = nov,
       volume = {795},
       number = {2},
          eid = {165},
        pages = {165},
          doi = {10.1088/0004-637X/795/2/165},
archivePrefix = {arXiv},
       eprint = {1405.5473},
 primaryClass = {astro-ph.GA},
       adsurl = {https://ui.adsabs.harvard.edu/abs/2014ApJ...795..165S}
}

@ARTICLE{2014A&A...563A..58T,
       author = {{Troncoso}, P. and {Maiolino}, R. and {Sommariva}, V. and {Cresci}, G. and {Mannucci}, F. and {Marconi}, A. and {Meneghetti}, M. and {Grazian}, A. and {Cimatti}, A. and {Fontana}, A. and {Nagao}, T. and {Pentericci}, L.},
        title = "{Metallicity evolution, metallicity gradients, and gas fractions at z \raisebox{-0.5ex}\textasciitilde 3.4}",
      journal = {\aap},
         year = 2014,
        month = mar,
       volume = {563},
          eid = {A58},
        pages = {A58},
          doi = {10.1051/0004-6361/201322099},
archivePrefix = {arXiv},
       eprint = {1311.4576},
 primaryClass = {astro-ph.CO},
       adsurl = {https://ui.adsabs.harvard.edu/abs/2014A&A...563A..58T}
}

@ARTICLE{2016ApJ...826L..11K,
       author = {{Kacprzak}, Glenn G. and {van de Voort}, Freeke and {Glazebrook}, Karl and {Tran}, Kim-Vy H. and {Yuan}, Tiantian and {Nanayakkara}, Themiya and {Allen}, Rebecca J. and {Alcorn}, Leo and {Cowley}, Michael and {Labb{\'e}}, Ivo and {Spitler}, Lee and {Straatman}, Caroline and {Tomczak}, Adam},
        title = "{Cold-mode Accretion: Driving the Fundamental Mass-Metallicity Relation at z \raisebox{-0.5ex}\textasciitilde 2}",
      journal = {\apjl},
         year = 2016,
        month = jul,
       volume = {826},
       number = {1},
          eid = {L11},
        pages = {L11},
          doi = {10.3847/2041-8205/826/1/L11},
archivePrefix = {arXiv},
       eprint = {1607.00014},
 primaryClass = {astro-ph.GA},
       adsurl = {https://ui.adsabs.harvard.edu/abs/2016ApJ...826L..11K}
}

@ARTICLE{2015ApJ...802L..26K,
       author = {{Kacprzak}, Glenn G. and {Yuan}, Tiantian and {Nanayakkara}, Themiya and {Kobayashi}, Chiaki and {Tran}, Kim-Vy H. and {Kewley}, Lisa J. and {Glazebrook}, Karl and {Spitler}, Lee and {Taylor}, Philip and {Cowley}, Michael and {Labbe}, Ivo and {Straatman}, Caroline and {Tomczak}, Adam},
        title = "{The Absence of an Environmental Dependence in the Mass-Metallicity Relation at z = 2}",
      journal = {\apjl},
         year = 2015,
        month = apr,
       volume = {802},
       number = {2},
          eid = {L26},
        pages = {L26},
          doi = {10.1088/2041-8205/802/2/L26},
archivePrefix = {arXiv},
       eprint = {1503.05559},
 primaryClass = {astro-ph.GA},
       adsurl = {https://ui.adsabs.harvard.edu/abs/2015ApJ...802L..26K}
}

@ARTICLE{2016MNRAS.463.2002H,
       author = {{Hunt}, Leslie and {Dayal}, Pratika and {Magrini}, Laura and {Ferrara}, Andrea},
        title = "{Coevolution of metallicity and star formation in galaxies to z ≃ 3.7 - I. A Fundamental Plane}",
      journal = {\mnras},
         year = 2016,
        month = dec,
       volume = {463},
       number = {2},
        pages = {2002-2019},
          doi = {10.1093/mnras/stw1993},
archivePrefix = {arXiv},
       eprint = {1608.05417},
 primaryClass = {astro-ph.GA},
       adsurl = {https://ui.adsabs.harvard.edu/abs/2016MNRAS.463.2002H}
}

@ARTICLE{2016ApJ...822...42O,
       author = {{Onodera}, M. and {Carollo}, C.~M. and {Lilly}, S. and {Renzini}, A. and {Arimoto}, N. and {Capak}, P. and {Daddi}, E. and {Scoville}, N. and {Tacchella}, S. and {Tatehora}, S. and {Zamorani}, G.},
        title = "{ISM Excitation and Metallicity of Star-forming Galaxies at z ≃ 3.3 from Near-IR Spectroscopy}",
      journal = {\apj},
         year = 2016,
        month = may,
       volume = {822},
       number = {1},
          eid = {42},
        pages = {42},
          doi = {10.3847/0004-637X/822/1/42},
archivePrefix = {arXiv},
       eprint = {1602.02779},
 primaryClass = {astro-ph.GA},
       adsurl = {https://ui.adsabs.harvard.edu/abs/2016ApJ...822...42O}
}

@ARTICLE{2017ApJ...849...39S,
       author = {{Suzuki}, Tomoko L. and {Kodama}, Tadayuki and {Onodera}, Masato and {Shimakawa}, Rhythm and {Hayashi}, Masao and {Tadaki}, Ken-ichi and {Koyama}, Yusei and {Tanaka}, Ichi and {Sobral}, David and {Smail}, Ian and {Best}, Philip N. and {Khostovan}, Ali A. and {Minowa}, Yosuke and {Yamamoto}, Moegi},
        title = "{The Interstellar Medium in [O III]-selected Star-forming Galaxies at z {\ensuremath{\sim}} 3.2}",
      journal = {\apj},
         year = 2017,
        month = nov,
       volume = {849},
       number = {1},
          eid = {39},
        pages = {39},
          doi = {10.3847/1538-4357/aa8df3},
archivePrefix = {arXiv},
       eprint = {1709.06731},
 primaryClass = {astro-ph.GA},
       adsurl = {https://ui.adsabs.harvard.edu/abs/2017ApJ...849...39S}
}
\bibliographystyle{aasjournal}

\clearpage
\appendix

\section{Emission Line Detections} \label{app: detections}

\begin{figure*}[h]
    \centering
    \includegraphics[width=18cm]{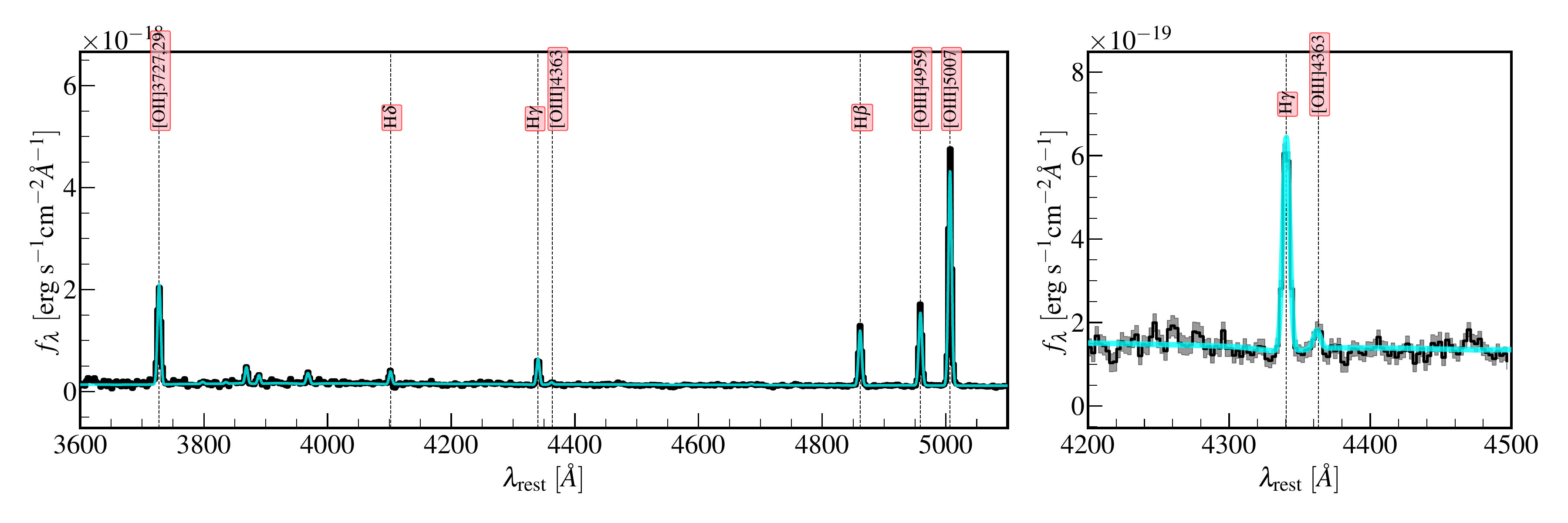}
    \caption{\textbf{[Left]} JWST/NIRSpec medium-resolution spectrum (G140M grating, gray) and best-fit pPXF model (cyan) for the 1181-00031514 galaxy at $z = 1.49$. \textbf{[Right]} Close-up view of the \oiii\ detection.}
    \label{fig: O4363_1181_00031514_g140m}
\end{figure*}

\begin{figure*}[h]
    \centering
    \includegraphics[width=18cm]{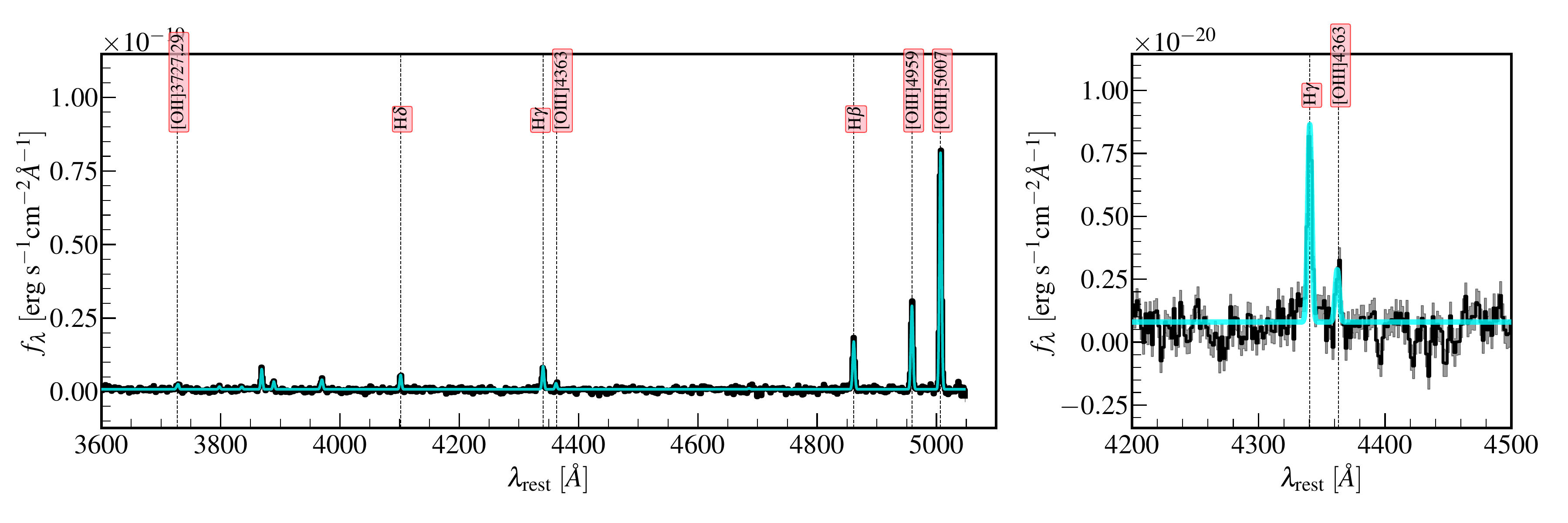}
    \caption{\textbf{[Left]} JWST/NIRSpec medium-resolution spectrum (G395M grating, gray) and best-fit pPXF model (cyan) for the 3215-00265801 galaxy at $z = 9.43$. This is the highest redshift entry in our calibration sample, and was originally reported in \cite{laseter+2024}. \textbf{[Right]} Close-up view of the \oiii\ detection.}
    \label{fig: O4363_3215_00265801_g395m}
\end{figure*}

\begin{figure*}[h]
    \centering
    \includegraphics[width=18cm]{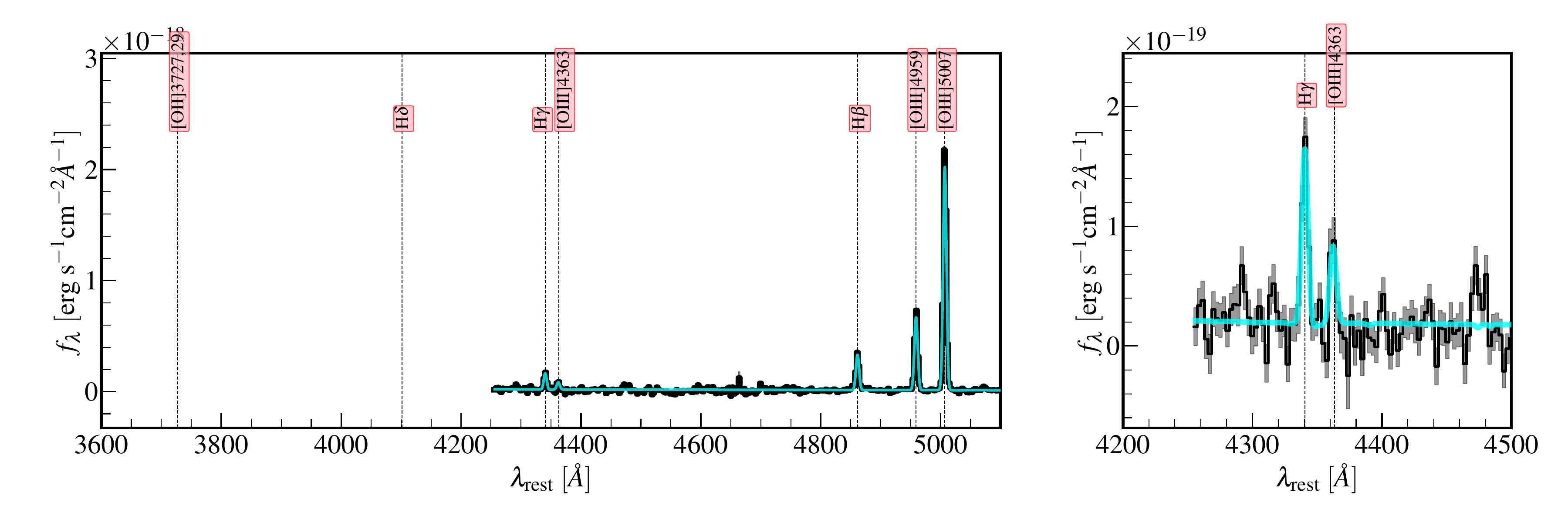}
    \caption{\textbf{[Left]} JWST/NIRSpec medium-resolution spectrum (G235M grating, gray) and best-fit pPXF model (cyan) for the 1181-00033391 galaxy at $z = 2.90$. \textbf{[Right]} Close-up view of the \oiii\ detection.}
    \label{fig: O4363_1181_00033391_g235m}
\end{figure*}

\begin{figure*}[h]
    \centering
    \includegraphics[width=18cm]{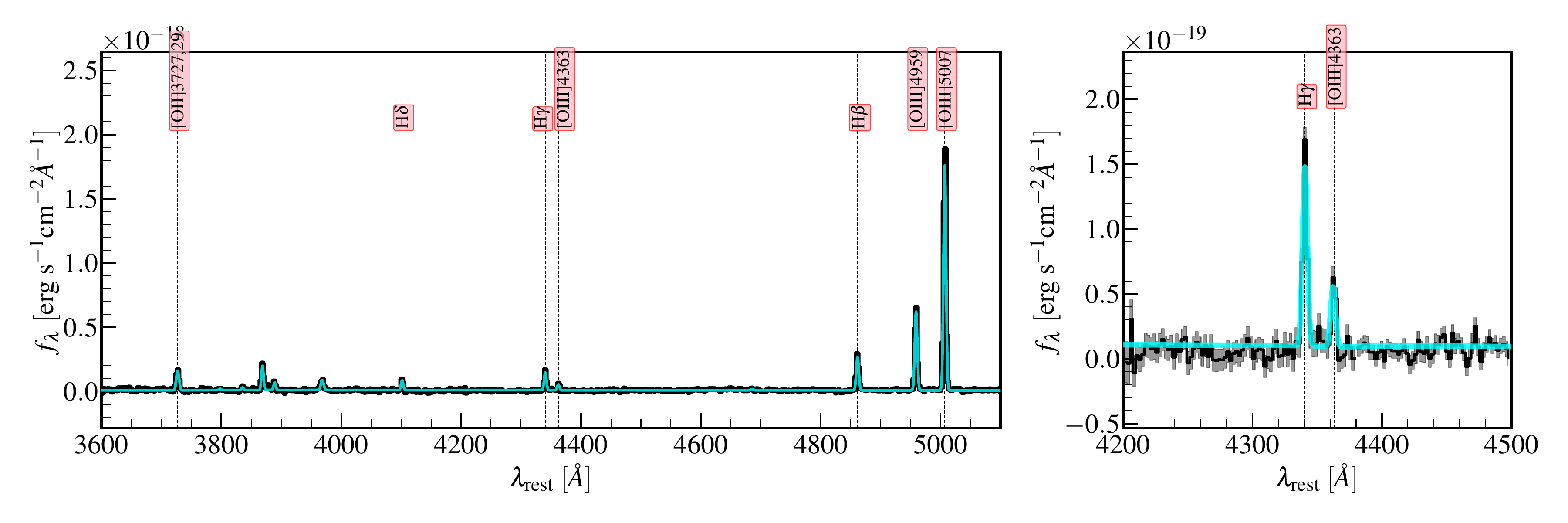}
    \caption{\textbf{[Left]} JWST/NIRSpec medium-resolution spectrum (G140M grating, gray) and best-fit pPXF model (cyan) for the 3215-00098554 galaxy at $z = 1.90$. \textbf{[Right]} Close-up view of the \oiii\ detection.}
    \label{fig: O4363_3215_00098554_g140m}
\end{figure*}

\begin{figure*}[h]
    \centering
    \includegraphics[width=18cm]{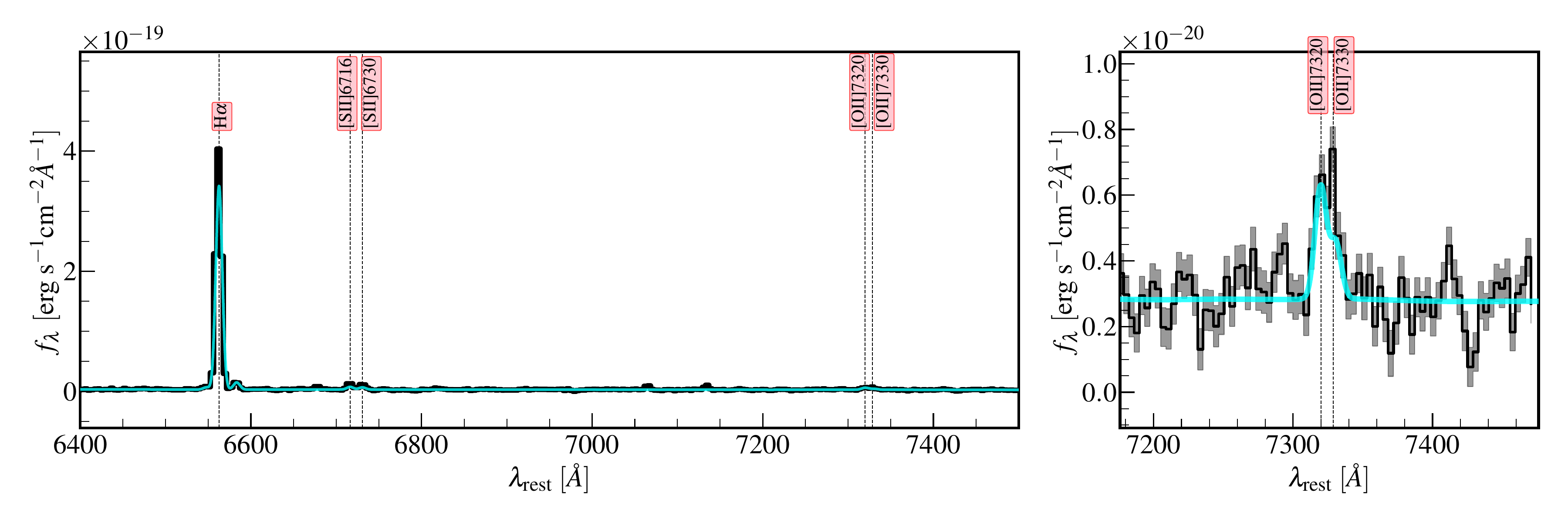}
    \caption{\textbf{[Left]} JWST/NIRSpec medium-resolution spectrum (G395M grating, gray) and best-fit pPXF model (cyan) for the 1180-00013596 galaxy at $z = 3.76$. \textbf{[Right]} Close-up view of the \oiidh\ detection.}
    \label{fig: O7320_1180_00013596_g395m}
\end{figure*}

\begin{figure*}[h]
    \centering
    \includegraphics[width=18cm]{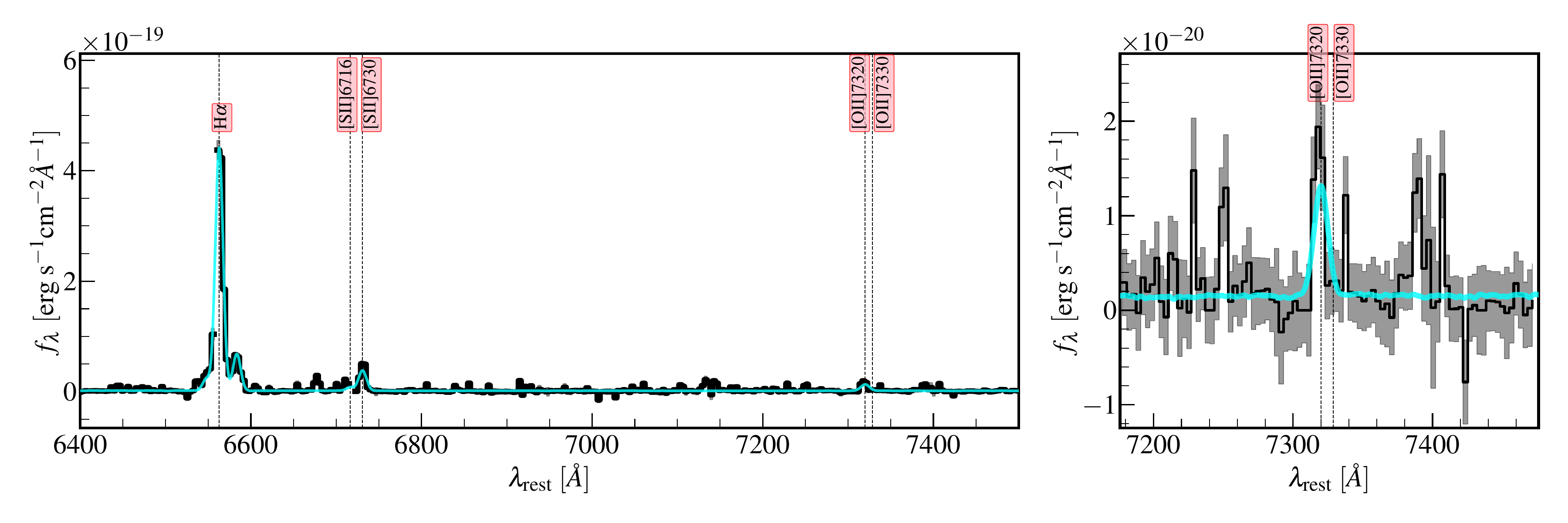}
    \caption{\textbf{[Left]} JWST/NIRSpec medium-resolution spectrum (G395M grating, gray) and best-fit pPXF model (cyan) for the 1180-00016375 galaxy at $z = 4.44$. \textbf{[Right]} Close-up view of the \oiidh\ detection.}
    \label{fig: O7320_1180_00016375_g395m}
\end{figure*}

\begin{figure*}[h]
    \centering
    \includegraphics[width=18cm]{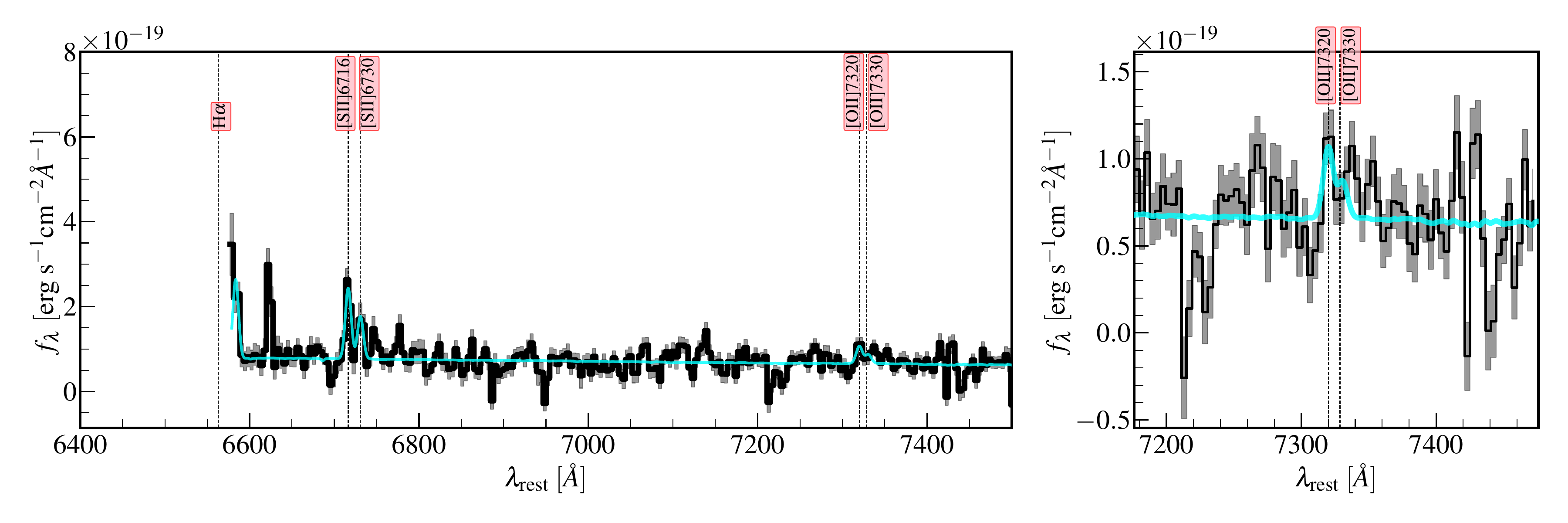}
    \caption{\textbf{[Left]} JWST/NIRSpec medium-resolution spectrum (G235M grating, gray) and best-fit pPXF model (cyan) for the 1181-00025030 galaxy at $z = 1.75$. \textbf{[Right]} Close-up view of the \oiidh\ detection.}
    \label{fig: O7320_1181_00025030_g235m}
\end{figure*}

\begin{figure*}[h]
    \centering
    \includegraphics[width=18cm]{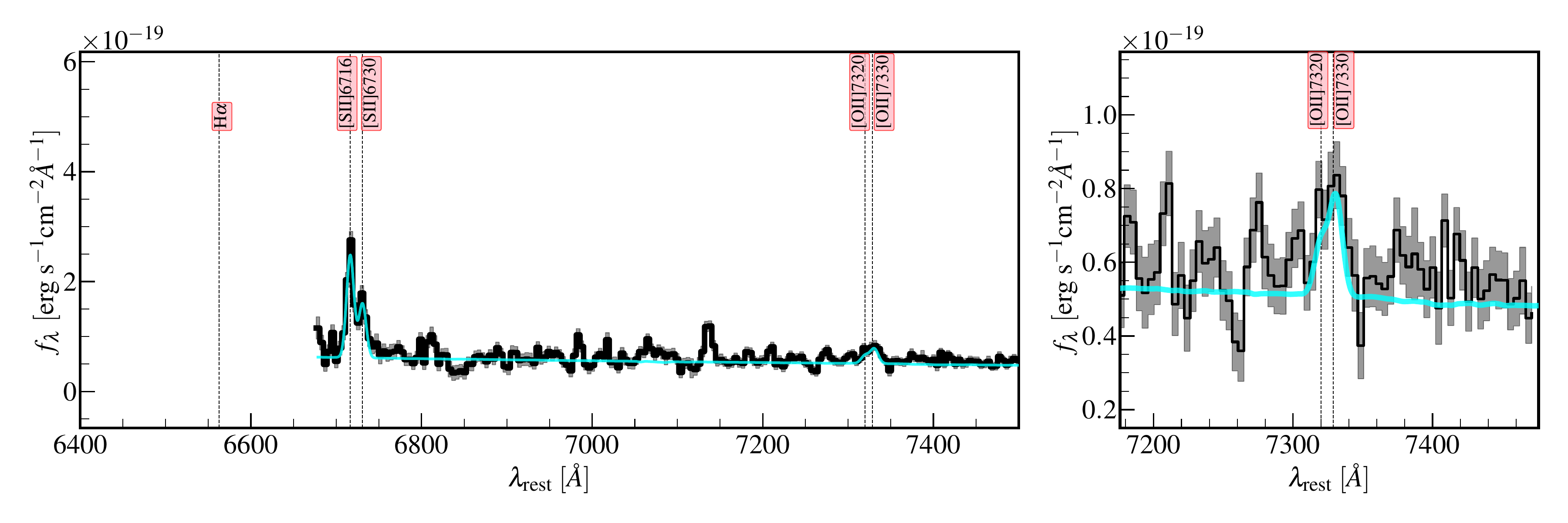}
    \caption{\textbf{[Left]} JWST/NIRSpec medium-resolution spectrum (G235M grating, gray) and best-fit pPXF model (cyan) for the 1181-00031514 galaxy at $z = 1.49$. \textbf{[Right]} Close-up view of the \oiidh\ detection.}
    \label{fig: O7320_1181_00031514_g235m}
\end{figure*}

Here, we present examples of our \oiii\ and \oiidh\ emission line detections in JWST/NIRSpec medium-resolution spectroscopy. Figures \ref{fig: O4363_1181_00031514_g140m}, \ref{fig: O4363_3215_00265801_g395m}, \ref{fig: O4363_1181_00033391_g235m}, and \ref{fig: O4363_3215_00098554_g140m} present \oiii\ detections. We note that the 3215-00265801 galaxy at $z = 9.43$ is the highest redshift entry in our calibration sample, originally discovered by \cite{laseter+2024}. Figures \ref{fig: O7320_1180_00013596_g395m}, \ref{fig: O7320_1180_00016375_g395m}, \ref{fig: O7320_1181_00025030_g235m}, and \ref{fig: O7320_1181_00031514_g235m} present \oiidh\ detections. 

\clearpage
\section{Literature Parametric Calibrations}

\begin{figure*}[h]
    \centering
    \includegraphics[width=16.5cm]{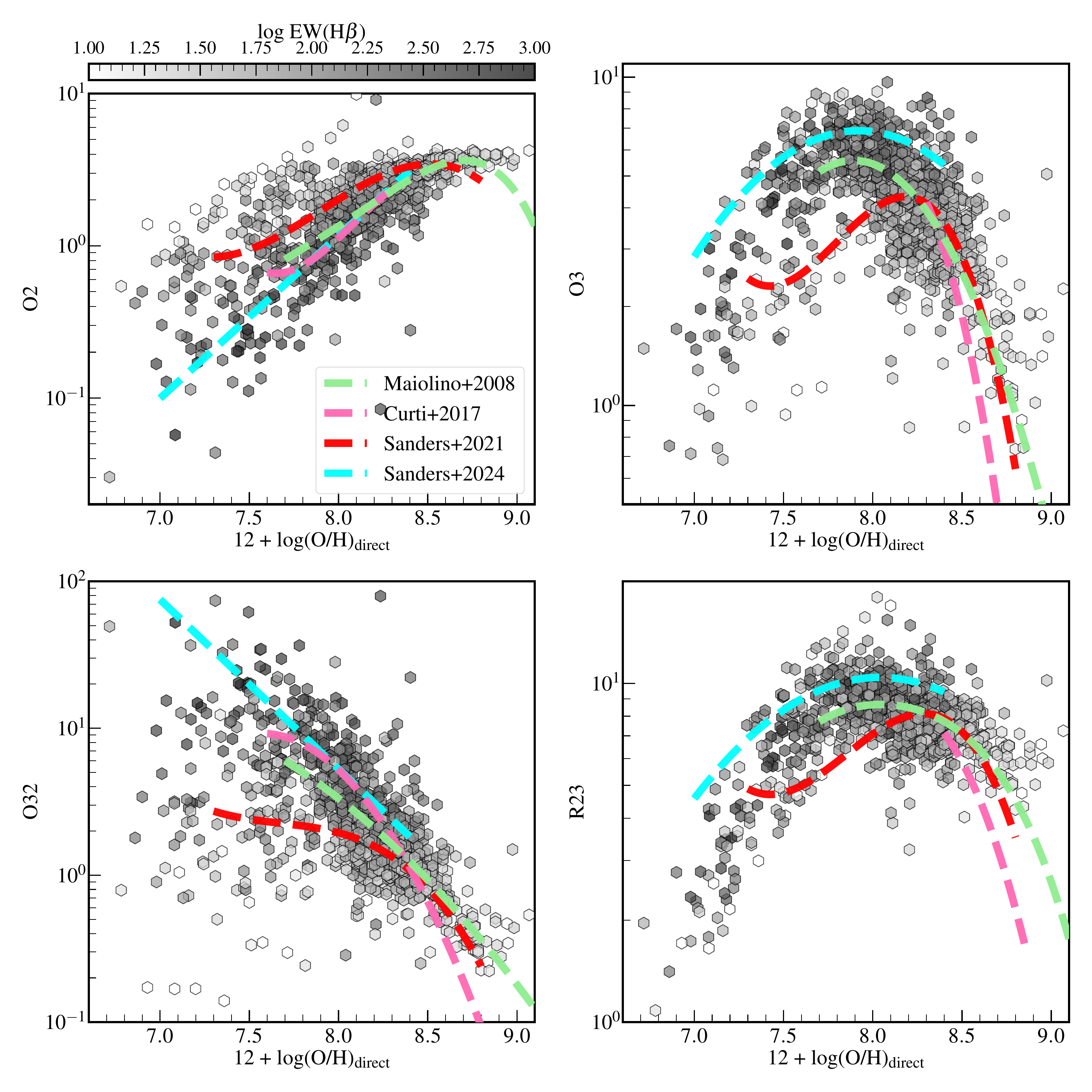}
    \caption{Comparison of some widely adopted parametric strong-line calibrations with our calibration data. The gray data points show the measurements for the galaxies in our calibration sample, color-coded with their corresponding \EWHb. The (dashed) green, pink, red, and cyan lines correspond to the best-fit polynomials from \cite{2008A&A...488..463M}, \cite{curti+2017}, \cite{sanders+2021}, and \cite{sanders+2023}, respectively. As shown, these polynomials cannot fully capture the complexities of the higher-order parameter space. This is evident from i) the large discrepancies in the best-fit polynomials reported in different studies; and ii) the large scatter of the calibration data around these best-fit polynomials.}
    \label{fig: calibration diagrams parametric}
\end{figure*}

Parametric strong-line calibrations often rely on polynomial fits to projections of calibration data in line ratio vs. metallicity 2D planes. As discussed in Section \ref{sec: intro}, these calibrations risk overlooking the complexities of the higher-order parameter space. Most often, even the 2D projections cannot be fully captured by polynomials. This is evident from the large scatter of calibration data around the best-fit polynomials in 2D projection planes. This has resulted in large discrepancies between different strong-line calibrations reported in the literature. We show this in Figure \ref{fig: calibration diagrams parametric}, where some widely adopted parametric strong-line calibrations are compared with our calibration data.

\clearpage
\section{Multivariate KDE implementation} \label{app: KDE}

In this Section, we discuss the details of our multivariate KDE implementations presented in Sections \ref{sec: direct: temperatures} and \ref{sec: strong}. Our KDE bandwidth selection is described in Section \ref{app: KDE-bandwidths}, while reliability tests beyond those presented in the main text are provided in Section \ref{app: KDE-validity}.

\subsection{Bandwidths} \label{app: KDE-bandwidths}

In this work, we adopt Gaussian kernels with bandwidths selected through a cross-validation algorithm. Bandwidth selection is performed using the \texttt{statsmodels} library \citep{seabold2010statsmodels}, adopting the maximum likelihood cross-validation approach suggested in \cite{9fb0f64c7113496fa27f2c3328ce9b97} which was originally proposed in \cite{Duin1976OnTC}. The selected bandwidths are then imported into our \texttt{scipy} implementation of KDE, described in Sections \ref{sec: direct: temperatures} and \ref{sec: strong}. The selected bandwidths for our \toii\ electron temperature estimator (Section \ref{sec: direct: temperatures}) in the 5-dimensional space of O2, O3, \EWHb, \toii, and \toiii\ are reported in Table \ref{table: bw-temp}. Moreover, the selected bandwidths for our gas-phase metallicity estimator (Section \ref{sec: strong}) in the 4-dimensional space of O2, O3, \EWHb, and gas-phase metallicity are reported in Table \ref{table: bw-met}. 

It is established that multivariate density estimates are far more sensitive to bandwidth selection than to the adopted kernel shape \citep{1986desd.book.....S, 1992mde..book.....S}. Here, we also explore how sensitive the accuracy of our electron temperature and metallicity estimators is to bandwidth selection. For this purpose, we adopt alternative bandwidths selected using the \texttt{scipy} implementation of Scott's rule of thumb \citep{1992mde..book.....S}: while the selected bandwidths become generally larger by $\sim 30\%$, the accuracy of both our electron temperature and gas-phase metallicity estimators remain intact. 

\subsection{Validation} \label{app: KDE-validity}

As the dimensionality of the parameter space increases, multivariate KDEs should be used with caution because the curse of dimensionality leads to an exponential growth in the sample size required to maintain constant accuracy \citep{1992mde..book.....S, DiMarzio, 2015arXiv150303305N}. This stems from an exponential increase in the variance term as the number of KDE covariates grows. Therefore, the addition of each new covariate must be carefully weighed against the trade-off between information gain and the variance inflation induced by the curse of dimensionality. As discussed in Sections \ref{sec: direct: temperatures} and \ref{sec: strong}, by removing the \EWHb\ covariate we construct lower-dimensional KDEs for both our electron temperature and gas-phase metallicity estimators, finding a decline in the accuracy of both. This indicates that by including the \EWHb\ axis we are winning the trade-off between information gain and variance inflation in both cases.

\begin{figure*}[h]
    \centering
    \includegraphics[width=13.75cm]{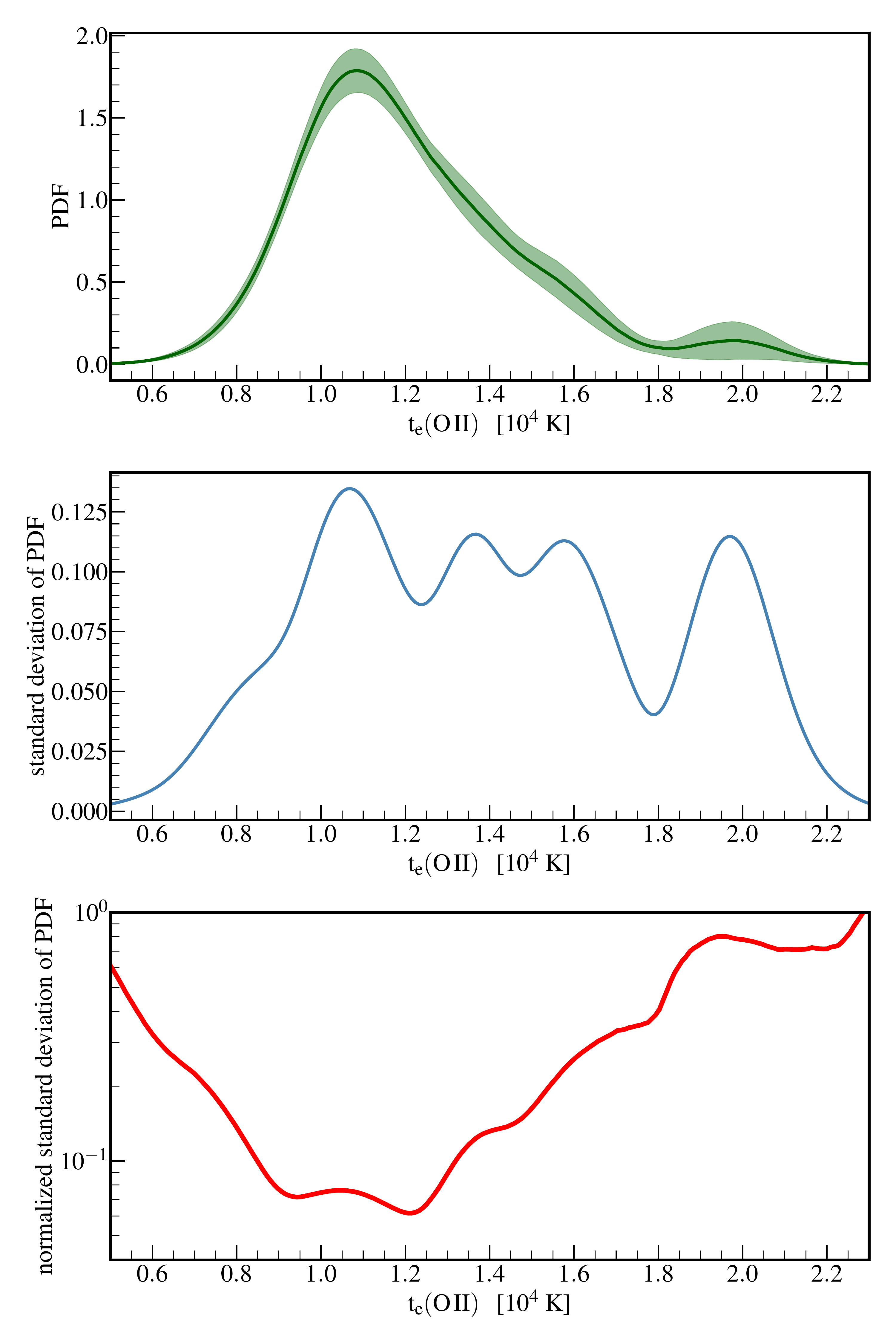}
    \caption{\toii\ dependence of the median bootstrapped PDF \textbf{[top]}, its standard deviation \textbf{[middle]}, and its normalized standard deviation \textbf{[bottom]} for our 5-dimensional electron temperature estimator on a median O2, O3, \EWHb, and \toiii\ slice. The bottom panel shows that the normalized scatter remains well below unity at across most of the \toii\ range, indicating that the KDE is stable. The normalized scatter increases noticeably at the edges of the electron temperature parameter space, where the calibration sample provides limited coverage.}
    \label{fig: bootstrap-temp-1D}
\end{figure*}

\begin{figure*}[h]
    \centering
    \includegraphics[width=13.75cm]{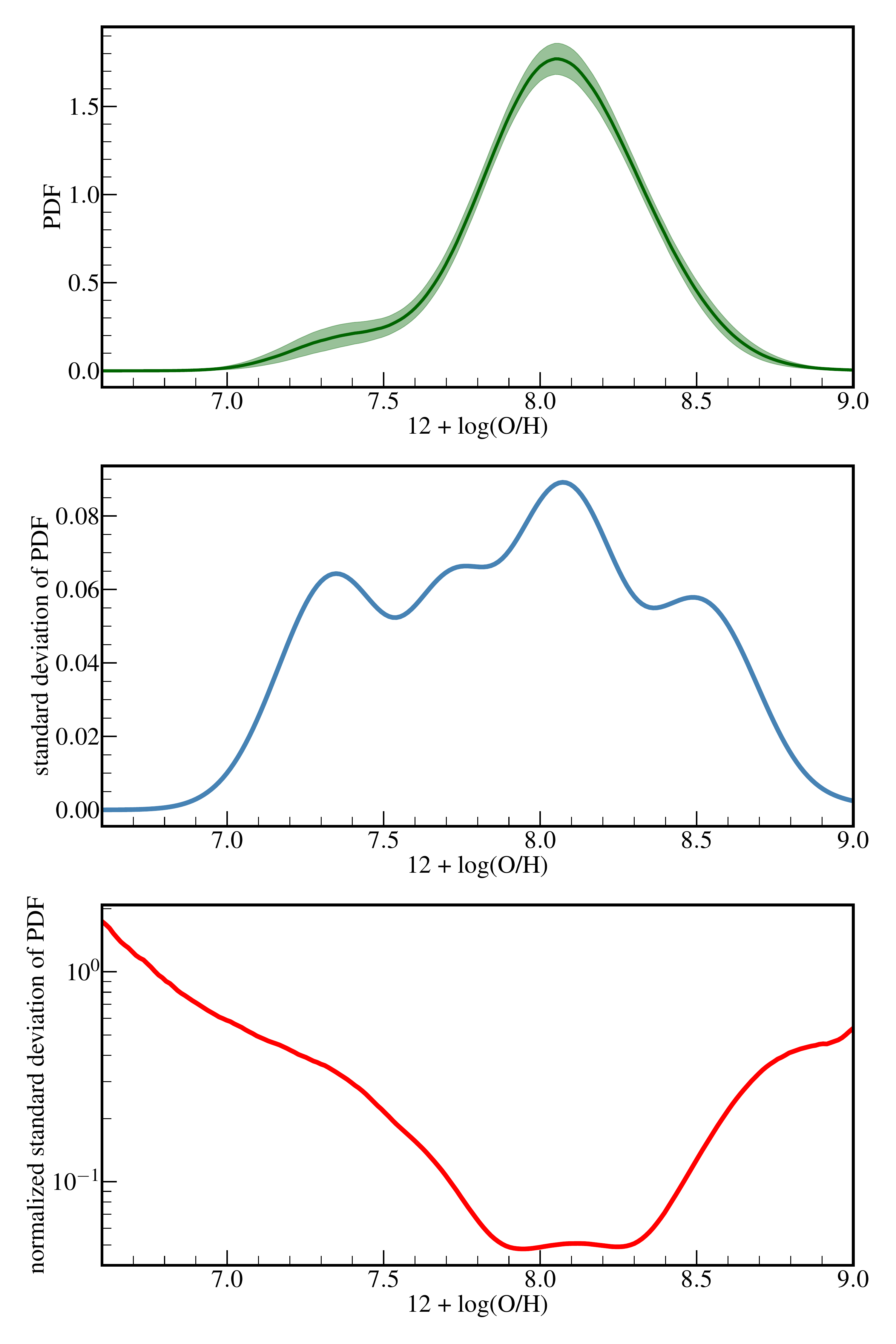}
    \caption{Gas-phase metallicity dependence of the median bootstrapped PDF \textbf{[top]}, its standard deviation \textbf{[middle]}, and its normalized standard deviation \textbf{[bottom]} for our 4-dimensional gas-phase metallicity estimator on a median O2, O3, and \EWHb\ slice. The bottom panel shows that the normalized scatter remains well below unity at across most of the gas-phase metallicity range, indicating that the KDE is stable. The normalized scatter increases noticeably at the edges of the gas-phase metallicity parameter space, where the calibration sample provides limited coverage.}
    \label{fig: bootstrap-met-1D}
\end{figure*}

We further assess the reliability of our density estimates through bootstrap resampling. We generate 1000 bootstrap realizations of the calibration sample and recompute the KDEs for each realization. To evaluate the stability of our density estimates, we examine the variation of PDFs estimated in these realizations around their median values. The three panels in Figure \ref{fig: bootstrap-temp-1D} show the \toii\ dependence of median PDF, its standard deviation, and its normalized standard deviation (defined as the standard deviation divided by the median value) for our 5-dimensional electron temperature estimator on a median O2, O3, \EWHb, and \toiii\ slice. Similarly, the three panels in Figure \ref{fig: bootstrap-met-1D} show the gas-phase metallicity dependence of median PDF, its standard deviation, and its normalized standard deviation for our 4-dimensional gas-phase metallicity estimator on a median O2, O3, and \EWHb\ slice. For both estimators the normalized scatter remains well below unity across most of the parameter range, indicating that the KDEs are stable. The only noticeable increase in normalized scatter occurs at the edges of the electron temperature and gas-phase metallicity parameter spaces, where the calibration sample provides limited coverage. 

\begin{figure*}[h]
    \centering
    \includegraphics[width=14cm]{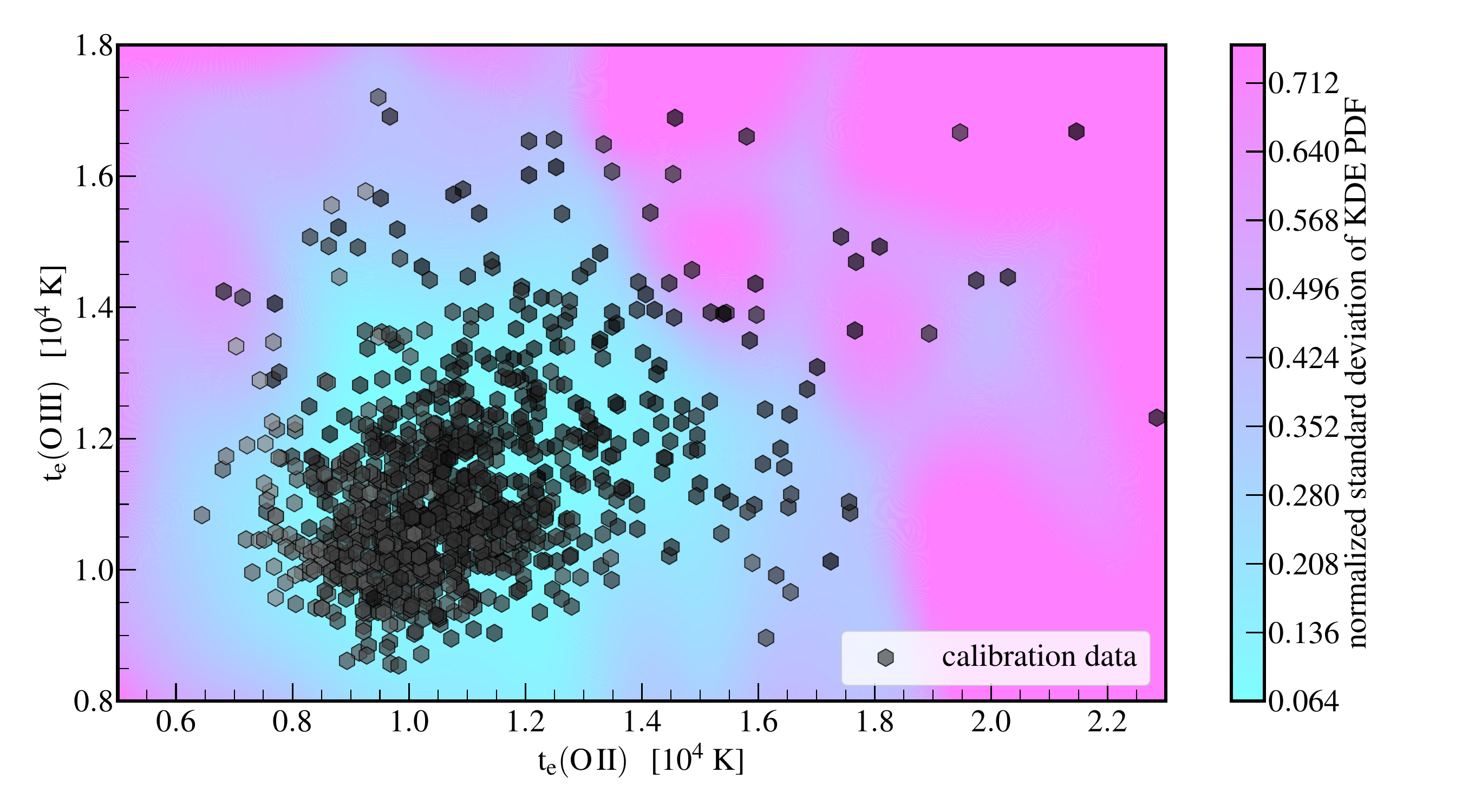}
    \caption{Normalized standard deviation of the bootstrapped PDF for our 5-dimensional electron temperature estimator in the 2-dimensional plane of median O2, O3, and \EWHb. Our density estimation is stable inside the region well covered by our calibration sample (small data points), as indicated by a normalized scatter well below unity.}
    \label{fig: bootstrap-temp-2D}
\end{figure*}

\begin{figure*}[h]
    \centering
    \includegraphics[width=14cm]{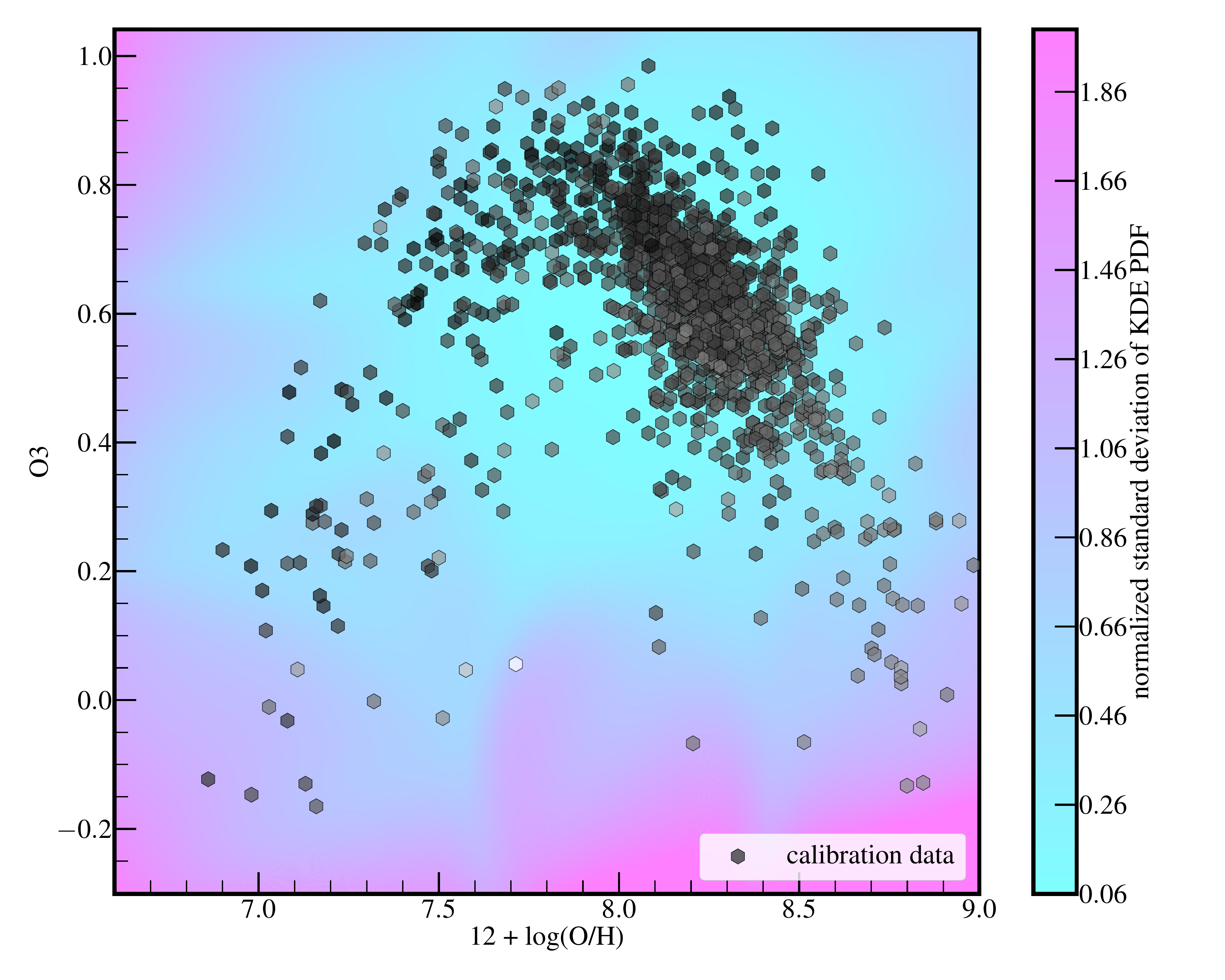}
    \caption{Normalized standard deviation of the bootstrapped PDF for our 5-dimensional gas-phase metallicity estimator in the 2-dimensional plane of median O2 and \EWHb. Our density estimation is stable inside the region well covered by our calibration sample (small data points), as indicated by a normalized scatter well below unity.}
    \label{fig: bootstrap-met-2D}
\end{figure*}

Similarly, Figure \ref{fig: bootstrap-temp-2D} shows the normalized standard deviation of the bootstrapped PDF for our 5-dimensional electron temperature estimator on a 2-dimensional plane of median O2, O3, and \EWHb. Figure \ref{fig: bootstrap-met-2D} shows the normalized standard deviation of the bootstrapped PDF for our 4-dimensional gas-phase metallicity estimator on a 2-dimensional plane of median O2 and \EWHb. Similar to Figures \ref{fig: bootstrap-temp-1D} and \ref{fig: bootstrap-met-1D}, these Figures indicate that our density estimations are stable inside the region well covered by our calibration sample. 

\clearpage

\clearpage

\begin{deluxetable}{lc}
\tablewidth{0pt}
\tablecaption{Selected bandwidths for our \toii\ electron temperature estimator (Section \ref{sec: direct: temperatures}) in the 5-dimensional space of O2, O3, \EWHb, \toii, and \toiii.}
\label{table: bw-temp}
\tablehead{
\colhead{parameter} & 
\colhead{bandwidth}
}
\startdata
$\log$(O2)        &  0.05 \\
$\log$(O3)        &  0.05 \\
$\log$(\EWHb)     &  0.11 \\
\toiii\;[$10^4$K] &  0.07 \\
\toii\;[$10^4$K]  &  0.09 \\
\enddata
\end{deluxetable}

\begin{deluxetable}{lc}
\tablewidth{0pt}
\tablecaption{Selected bandwidths for our gas-phase metallicity estimator (Section \ref{sec: strong}) in the 4-dimensional space of O2, O3, \EWHb, and gas-phase metallicity.}
\label{table: bw-met}
\tablehead{
\colhead{parameter} & 
\colhead{bandwidth}
}
\startdata
$\log$(O2)     &  0.05 \\
$\log$(O3)     &  0.05 \\
$\log$(\EWHb)  &  0.14 \\
$\metallicity$ &  0.09 \\
\enddata
\end{deluxetable}

\end{document}